\documentclass{cernyrep} 
\usepackage{texnames}
\usepackage[T1]{fontenc}
\usepackage[figuresright]{rotating}
\newcommand{\fNTq}[1]{\ensuremath{f_{T_{#1}}^{(N)}}}

\newcommand{\BNq}[1]{\ensuremath{B_{#1}^{(N)}}}

\newcommand{\SigmapiN}{\ensuremath{\Sigma_{\pi\!{\scriptscriptstyle N}}}}
\def\ga{\mathrel{\raise.3ex\hbox{$>$\kern-.75em\lower1ex\hbox{$\sim$}}}}
\def\la{\mathrel{\raise.3ex\hbox{$<$\kern-.75em\lower1ex\hbox{$\sim$}}}}

\begin{document}

\def\beq{\begin{equation}}
\def\eeq{\end{equation}}
\def\la{~\mbox{\raisebox{-.6ex}{$\stackrel{<}{\sim}$}}~}
\def\ga{~\mbox{\raisebox{-.6ex}{$\stackrel{>}{\sim}$}}~}
\def\iso#1#2{\mbox{${}^{#2}{\rm #1}$}}
\def\h#1{\iso{H}{#1}}
\def\he#1{\iso{He}{#1}}
\def\li#1{\iso{Li}{#1}}
\def\be#1{\iso{Be}{#1}}
\def\b#1{\iso{B}{#1}}
\def\bor#1{\iso{B}{#1}}

\title{The violent Universe: the Big Bang\footnote{Set of four lectures given at the 
2009 European School of High-Energy Physics, Bautzen, Germany, June 2009.}}
 
\author{K.~A.~Olive}

\institute{William I. Fine Theoretical Physics Institute, University of Minnesota, Minneapolis, MN 55455, USA}

\maketitle 

\begin{abstract}
\vskip - 2.2in
\rightline{UMN--TH--2832/10}
\rightline{FTPI--MINN--10/03}
\rightline{January 2010}
\vskip 1.4in
Four lectures on Big Bang cosmology, including microwave background
radiation, Big Bang nucleosynthesis, dark matter, inflation, and baryogenesis. 
\end{abstract}

The Big Bang theory provides a detailed description of the history and evolution of the 
Universe. Direct experimental and observational evidence allows us to probe back
to the first second after the initial state (bang) when the temperatures were of order
1 MeV and the light elements were created.   Our understanding of the Standard Model
of electroweak interactions allows us to push the description of the early universe back to 
about $10^{-10}$ s  after the bang when we expect that the electroweak symmetry was restored. 
Indeed, it is possible to discuss events back to the Planck time ($10^{-44}$ s after the bang)
albeit in a very model dependent way.

In these four lectures, I hope to give an overview of modern cosmology
with an emphasis on particle interactions in cosmology.  
After a description of the standard FLRW model (including the microwave background radiation)
in Lecture 1, I will cover both inflation and baryogenesis in Lecture 2. Lecture 3 will 
focus on Big Bang Nucleosynthesis (BBN) and Lecture 4 on dark matter.

\section{Lecture 1: Standard Cosmology}

\subsection{The FLRW metric and its consequences}

The standard Big Bang model assumes homogeneity and
isotropy.  As a result,  one can construct the space-time metric
by embedding a maximally symmetric three dimensional space in a 
four dimensional space-time (see, e.g., Ref.~\cite{wein}). The most general form for a metric of this type is the 
Friedmann--Lema\^{i}tre--Robertson--Walker metric which in co-moving coordinates is given by
\beq
	ds^2  = dt^2  -  R^2(t)\left[ {dr^2 \over \left(1-kr^2\right) }
      + r^2 \left(d\theta^2  + \sin^2 \theta d\phi^2 \right)\right]	,
\label{met}
\eeq
where $R(t)$ is the cosmological scale factor and $k$ the three-space
curvature constant ($k = 0, +1, -1$ for a spatially flat, closed or open
universe). $k$ and $R$ are the only two quantities in the
metric which distinguish it from flat Minkowski space.
It is  also common to assume
 the perfect fluid form for the energy-momentum
tensor
\beq
	T^{\mu\nu}   = pg^{\mu\nu}   + (p + \rho)u^\mu u^\nu 		,
\eeq
where $g_{\mu\nu}$   is the space-time metric described by Eq.~(\ref{met}),
 $p$ is the isotropic
pressure, $\rho$ is the energy density and $u^\mu  = (1,0,0,0)$
 is the velocity vector
for the isotropic fluid.  The $(00)$ component of Einstein's equation  
\beq 
R_{\mu\nu} - \frac{1}{2}g_{\mu\nu}R - \Lambda g_{\mu\nu} = 8 \pi G_N T_{\mu\nu}
\eeq
yields the
Friedmann equation
\beq
	H^2  \equiv \left({\dot{R} \over R}\right)^2  = { 8 \pi G_N \rho \over 3}
 - { k \over R^2}  + {\Lambda \over 3} ,
\label{H}
\eeq
and the $(ii)$ components give
\beq
	\left({\ddot{R} \over R}\right) = {\Lambda \over 3} -
 { 4 \pi G_N ( \rho + 3p) \over 3} ,
 \label{ii}
\eeq
where $\Lambda$ is the cosmological constant.
In addition, from ${T^{\mu\nu}}_{;\nu}   =  0$, we obtain
\beq
	\dot{\rho} = -3H(\rho + p)	.
\label{rhod}	
\eeq
Note that of these last three equations, only two are actually independent.
These equations form the basis of the standard Big Bang model.

At early times ($t < 10^5 $ yrs) the Universe is thought to have been
dominated by radiation so that the equation of state can be given by $p =
\rho/3$.  If we neglect the contributions to $H$ from $k$ and $\Lambda$
 (this is always a
good approximation for small enough $R$) then we find that
\beq
	R(t) \sim t^{1/2} \qquad    
\eeq
and $\rho \sim R^{-4}$  so that $t \sim (3/32 \pi G_N\rho)^{1/2}$.
  Similarly for a matter or dust
dominated universe with $p = 0$,
\beq
	R(t) \sim t^{2/3}    
\eeq
and $\rho \sim R^{-3}$.  The Universe makes the transition
 between radiation and matter
domination when $\rho_{rad} = \rho_{matter}$ or 
when $T \simeq$ few $\times~10^3$ K.
In a vacuum or $\Lambda$ dominated universe (that we are approaching today)
\beq
R(t) \sim e^{\sqrt{\Lambda/3}\  t} .
\label{Vdom}
\eeq

\begin{figure}
\centering\includegraphics[width=.5\linewidth]{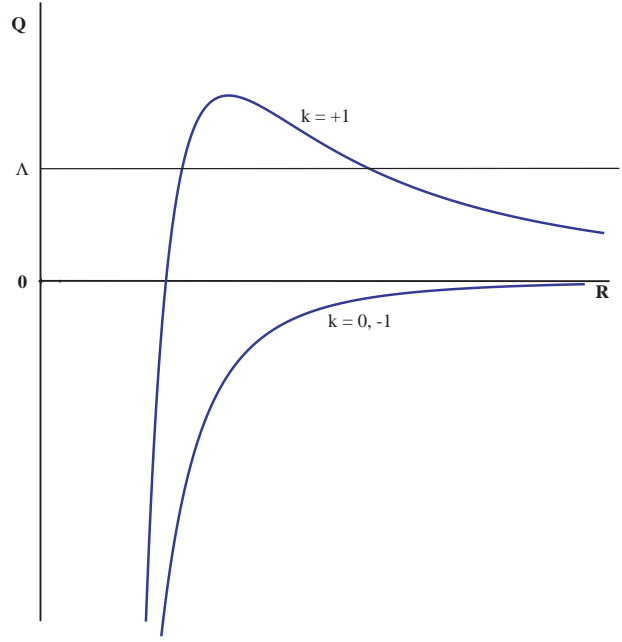}
\caption{The qualitative behaviour of the scale factor $R$ as obtained using the condition
$Q \le \Lambda$ for the three choices of the curvature constant $k$}
\label{fig:Q}
\end{figure}

More general solutions for the behaviour of the scale factor can easily be
found by defining a quantity $Q$:
\beq
Q = \frac{3k}{R^2} - 8\pi G_N \rho .
\eeq
If we further assume $p = (\gamma-1) \rho$ (with $\gamma$ between 1 and 2),
we have that $\rho \sim R^{-3\gamma}$ and we can write
\beq
\frac{\dot R}{R} = \pm \left[ \frac{\Lambda - Q}{3} \right]^{1/2} ,
\label{Qevol}
\eeq
implying that $Q \le \Lambda$.  
For all choices of $k$, the function $Q \to -\infty$ as $R \to 0$.
When $k = +1$, it is easy to see that $Q$  has  a maximum (with $Q > 0$) and 
then tends to 0 as $R \to \infty$. 
When $k = 0$ or $-1$, $Q$ monotonically increases towards 0 as $R \to \infty$.
In Fig.~\ref{fig:Q}, the function $Q$ is plotted qualitatively as a function of the scale factor.

Let us first consider the more interesting case of $k = +1$.
For a cosmological constant $\Lambda  > Q_{max}$, $Q< \Lambda$ for all $R$.
In this case, two distinct solutions are possible: the universe may expand forever from a singularity
at $R = 0$ to  $R = \infty$, or by choosing the lower sign in Eq.~(\ref{Qevol}),
the universe may collapse from infinity to a singularity.

There are also two possible solutions for $0 < \Lambda < Q_{max}$, the case depicted in Fig.~\ref{fig:Q}.
The universe may again start at a singularity at $R=0$ and expand to the point where 
$Q = \Lambda$ and then recollapse.  Alternatively, the Universe may start from infinity and collapse to the
point where $Q =  \Lambda$ (at larger $R$),   bounce back and expand to infinity.
When $\Lambda = Q_{max}$ there are a total of five solutions.
At the value of $R$ such that $Q = Q_{max}$, we have the Einstein Static Universe.
This is the only solution which is neither expanding nor contracting.  The remaining four solutions
either asymptotically expand or contract towards or away from the Einstein Static case. 
Finally for $\Lambda < 0$, there is only one solution for which the universe expands and subsequently 
recollapses. 

Finding the solutions when $k = 0$ or $-1$ is similar, and the behaviour of $R$ depends only on whether
$\Lambda$ is positive or negative. For $\Lambda \ge 0$ there are two solutions, 
for $\Lambda < 0$, there is only one. When $\Lambda = 0$, we obtain the standard notion that closed universes recollapse, while open and flat universes expand forever. When $\Lambda \ne 0$, these
simple associations are spoiled as a closed universe can expand forever (for large enough $\Lambda$)
and open and flat universes can recollapse (for $\Lambda < 0$).

Exact solutions to the equations of motion can be obtained relatively easily in terms of conformal 
coordinates. We can, for example, rewrite the metric as
\beq
ds^2  = dt^2  - R^2(t)\left[ {d\chi^2 }
      + f^2(\chi) \left(d\theta^2  + \sin^2 \theta d\phi^2 \right)\right]	 ,
\eeq
where 
\beq
f(\chi) = \left\{ \begin{array}{ll}
 \sinh \chi & k = -1\\
 \chi & k = 0 \\
\sin \chi & k = +1 \end{array} \right. .
\eeq
We can go further and define a conformal time coordinate using $R d\eta = dt$ so that 
\beq
ds^2  = R^2(\eta )\left[  dt^2  -  {d\chi^2 }
      + f^2(\chi) \left(d\theta^2  + \sin^2 \theta d\phi^2 \right)\right]	 .
\eeq
In terms of these coordinates, the Friedmann equation becomes
\beq
R^{\prime\prime} + kR =\frac{4 \pi G_N}{3} (\rho - 3p) R^3 ,
\eeq
where $^\prime$ denotes a derivative with respect to $\eta$ and is easily solved
\beq
R \propto \left\{ \begin{array}{ll}
 \cosh \eta -1 & k = -1\\
 \eta^2/2 & k = 0 \\
1 - \cos \eta & k = +1 \end{array} \right.  \qquad
t \propto  \left\{ \begin{array}{ll}
 \sinh \eta -\eta & k = -1\\
 \eta^3/6 & k = 0 \\
\eta - \sin \eta & k = +1 \end{array} \right. 
\eeq
for $p=0$ and
\beq
R \propto \left\{ \begin{array}{ll}
 \sinh \eta  & k = -1\\
 \eta & k = 0 \\
\sin \eta & k = +1 \end{array} \right.  \qquad
t \propto  \left\{ \begin{array}{ll}
 \cosh \eta -1 & k = -1\\
 \eta^2/2 & k = 0 \\
1 - \cos \eta & k = +1 \end{array} \right. 
\eeq
for $p= \rho/3$.
 
In the absence of a cosmological constant, one can 
define a critical energy density $\rho_c$
  such that $\rho =\rho_c$  for $k = 0$
\beq
	\rho_c  = 3H^2 / 8 \pi G_N		.
\eeq
In terms of the present value of the Hubble parameter this is
\beq
	\rho_c  = 1.88 \times 10^{-29} {h_0}^2  {\rm ~g~cm}^{-3}  ,
\eeq
where
\beq
	h_0  = H_0 /(100 {\rm ~km~Mpc}^{-1}   {\rm ~s}^{-1}  )		.
\eeq
The cosmological density parameter is then defined by
\beq
	\Omega \equiv {\rho \over \rho_c} 	.		
\eeq
It is useful to also define a deceleration parameter 
\beq 
q_0 = -\frac{{\ddot R}_0 R_0}{{\dot R}^2_0} .
\eeq
This standard definition was formulated under the presumption that the expansion rate
of the Universe is in fact slowing down.  As noted above, and discussed further below, 
modern measurements indicate the opposite. That is, the expansion
is accelerating (meaning that $q_0 < 0$). 
The $(ii)$ component, Eq.~(\ref{ii}), can be written in terms of $q_0$ as
\beq
-2 q_0 H_0^2 = \frac{2 \Lambda}{3} - \frac{8 \pi G_N \rho_0}{3}
\eeq
when the pressure is neglected.
This can be combined with the Friedmann equation, Eq.~(\ref{H}), and rewritten as
\beq
	 {k \over R_0^2} = \Lambda + H_0^2 (2q_0 - 1)	 ,
\label{o-1}	
\eeq
or
\beq
 {k \over R_0^2} =  H_0^2 (\frac{3}{2}\Omega_0 -q_0 - 1)	.
\eeq
Furthermore, when $\Lambda = 0$, $q_0 = \Omega_0/2$
so that $k = 0, +1, -1$ corresponds to $\Omega = 1, \Omega > 1$
 and $\Omega < 1$.  
Observational limits on $h_0$ and $\Omega$ are~\cite{wmap}
\beq
h_0 = 0.71 \pm 0.01  \qquad \Omega_0 = 1.006 \pm 0.006.
\label{range}
\eeq

\begin{figure}
\centering\includegraphics[width=.7\linewidth]{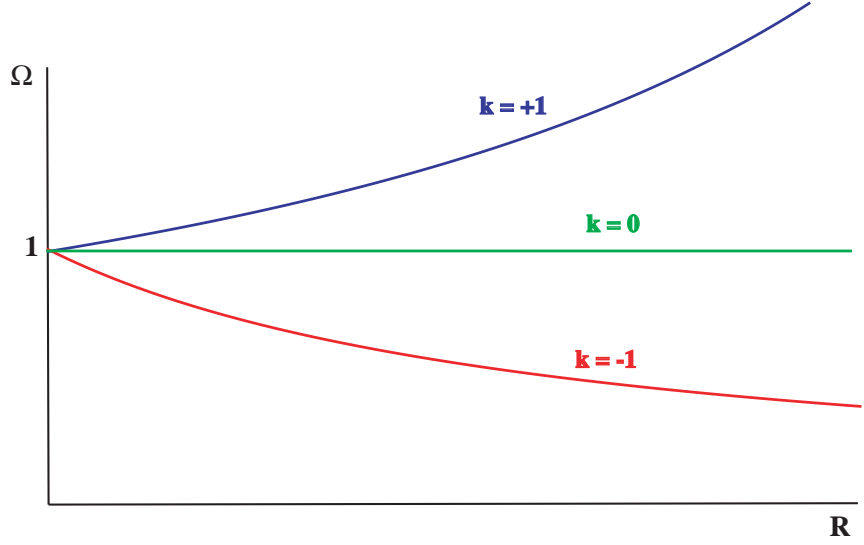}
\caption{The evolution of the cosmological density parameter $\Omega$
as a function of the scale factor for a closed, open and 
spatially flat Universe}
\label{fig:myfig}
\end{figure}

It is important to note that $\Omega$ is a function of 
time or of the scale factor.
The qualitative evolution of $\Omega$ is shown in Fig.~\ref{fig:myfig} for $\Lambda = 0$.
For a spatially flat Universe, $\Omega = 1$ always. When $k = +1$,
there is a maximum value for the scale factor $R$. At early
times (small values of $R$), $\Omega$ always tends to one.
Note that the fact that we do not yet know the sign of
$k$, or equivalently whether $\Omega$ is larger than or smaller than
unity, implies that we are at present still at the very left in the
figure.  What makes this peculiar is that one would normally expect
the sign of $k$ to become apparent after a Planck time of $10^{-44}$ s.
It is extremely puzzling that some $10^{60}$ Planck times later,
we still do not know the sign of $k$.

The Friedmann equation also lends itself to integration to determine the 
age of the Universe.  Note that for a given equation of state, we can write
$\rho = \rho_0 (R_0/R)^{3\gamma}$, and
\beq
H_0 t = \int_0^1 \frac{dx}{\left[ 1 - \Omega_0 - \Omega_\Lambda + \Omega_0 x^{2 - 3\gamma}+ \Omega_\Lambda x^2  \right]^{1/2} } ,
\eeq
where $\Omega _\Lambda = \Lambda/3 H_0^2$.
When $\Lambda = 0$ and $\Omega_0 = 1$, this is easily integrated to give
\beq 
t = \frac{2}{3 H_0}  \qquad \gamma = 1
\eeq
or 
\beq
t = \frac{1}{2 H_0}  \qquad \gamma = \frac{4}{3} .
\eeq

Because of the finite age of the Universe in the Big Bang model,
there is a particle horizon corresponding to the maximum distance 
traversed by light. In general, we can write the proper distances between 
a location specified by comoving coordinate $t_1, r_1$ and our location at $t$ and $r=0$ 
in terms of the metric~(\ref{met}) 
\beq
d_p = R(t) \int_0^{r_1} \frac{dr^\prime}{\sqrt{1 - k {r^\prime}^2}} .
\eeq
For light paths ($ds^2 = 0$)
\beq
\int_{t_1}^t \frac{dt^\prime}{R(t^\prime)} = \int_0^{r_1} \frac{dr^\prime}{\sqrt{1 - k {r^\prime}^2}} .
\eeq
As $t_1 \to 0$, $r_1$ becomes the maximum coordinate distance from which 
we can receive a signal.
Thus the particle horizon is defined by
\beq
d_H = R(t) \int_0^{r_H} \frac{dr^\prime}{\sqrt{1 - k {r^\prime}^2}} = R(t) \int_0^t \frac{dt^\prime}{R(t^\prime)} .
\label{dh}
\eeq
For $\Lambda = 0$ and $\Omega_0 = 1$, we again obtain very simple solutions:
\beq 
d_H  =3 t  \qquad \gamma = 1
\eeq
or 
\beq
d_H =2t  \qquad \gamma = \frac{4}{3} .
\eeq
Note that because $d_H/R$ grows with time, we see more of the Universe as time goes on.
That is, new information is continuously entering our particle horizon.

\subsection{The hot thermal Universe}

 The epoch of recombination occurs when electrons
 and protons form neutral hydrogen  through 
$e^{ -  } +  p     \rightarrow $  H  $+   \gamma $
  at a temperature  
$T_{ R} {~ \sim }$~few $\times 10^{3}$ K  ${~\sim}~1$ eV.  For $T < T_{R}$, 
photons are decoupled while for $T > T_{ R}$,  photons are
 in thermal equilibrium and at higher temperatures, the Universe is
radiation dominated and the content
 of the radiation plays a very important role.  Today, the content
 of the microwave background consists of photons with
$T_0 =  2.725  \pm 0.001$ K~\cite{cobem}.  
We can calculate the energy density of photons from
\beq
 \rho_\gamma  = \int E_\gamma dn_\gamma ,
\label{rhog}
\eeq
 where the density of states is given by
\beq
dn_\gamma  =   {g_\gamma \over 2 
 \pi^{ 2}}[exp(E_\gamma/T)-1]^{ -  1} q^{ 2} dq , 
\eeq
 and $g_\gamma = $  2 simply counts the number of degrees of freedom
 for photons,
$E_\gamma =  q$ is just the photon energy (momentum). 
 (I am using units such that  
$\hbar =  c  = k_{ B}   =$  1 and will do so through the remainder
 of these lectures.)  
Integrating Eq.~(\ref{rhog}) gives
\beq
\rho_\gamma = {\pi^2 \over 15} T^4 ,
\eeq
 which is the familiar blackbody result. In addition, we also have
 \beq
 p = \frac{1}{3} \rho \qquad s = \frac{4}{3} \frac{\rho}{T} \qquad n_\gamma = \frac{2 \zeta(3)}{\pi^2} T^3 .
 \eeq

In general, at very early times, at very high temperatures,
 other particle degrees of freedom join the radiation background when  
$T{~\sim }~m_{i}$  for each  particle type $i$ if that type is brought
 into thermal equilibrium through interactions.  In equilibrium, 
the energy density of a particle type $i$ is given by
\beq
 \rho_{i}  = \int E_{i} dn_{q_{i}} 
\eeq
 and
\beq 
 dn_{q_{i}} = {g_{i} \over 2  \pi^{ 2}}[\exp[(E_{q_{i}} - 
\mu_{i})/T] \pm 1]^{ -1 }q^{2}dq ,
\eeq
where again $g_{i}$ counts the total number of degrees of freedom for type $i$,
\beq
 E_{q_{i}} =  \left(m_{i}^{2} + q_{i}^{ 2}\right)^{1/2} ,
\eeq
$\mu_{i}$ is the chemical potential if present and  $ \pm$  
corresponds to either Fermi or Bose statistics.

In the limit that  $T \gg m_{i} $  the total energy density can
 be conveniently expressed by  
\beq
 \rho {} = \left( \sum_B g_{B} + {7 \over 8} \sum_F  g_{F} \right)
   {\pi^{ 2} \over 30}  T^{4}     \equiv    {\pi^{ 2} \over 30} N(T) T^{4} ,
\label{NT}
\eeq
 where $g_{B(F)} $  are the total number of boson (fermion) 
degrees of freedom and the sum runs over all boson (fermion) states with 
$m \ll T$.  The factor of 7/8 is due to the difference between
 the Fermi and Bose integrals.  Equation~(\ref{NT}) defines $N(T)$
 by taking into account  new particle degrees 
of freedom as the temperature is raised.  

In the radiation dominated epoch,
Eq.~(\ref{rhod}) can be integrated (neglecting the $T$-dependence of $N$)
giving us a relationship between
 the age of the Universe and its temperature
\beq
 t = \left({90 \over 32 \pi^3 G_{ N} N(T)}\right)^{ 1/2}  T^{ -  2} .
\eeq
 Put into a more convenient form
 \beq
 tT_{\rm MeV}^{ 2}  =
2.4 [N(T)]^{-1/2}     .
\label{tT2}
\eeq
 where $t$ is measured in seconds and
$T_{\rm MeV} $  in units of MeV.

 The value of $N(T)$ at any given temperature depends
 on the particle physics model.  In the standard $SU(3) \times
SU(2)\times U(1)$  model, we can specify $N(T)$ up to 
temperatures of order 100~GeV.
The value of $N$ in the Standard Model can be seen in Table~\ref{tab:NT}.

\begin{table}
\caption{Effective numbers of degrees of freedom in the Standard Model}
\begin{center}
\begin{tabular}{llc}
\hline
{\bf Temperature} & {\bf New particles} \qquad
&\boldmath$4N(T)$ \\
\hline\rule{0pt}{12pt}
$T < m_{ e}$   &     $\gamma$'s +   $\nu$'s & 29 \\
$m_{ e} <   T  < m_\mu$ &    $e^{\pm}$ & 43 \\
$m_\mu <  T  < m_\pi$  &   $\mu {}^{\pm}$ & 57 \\
$m_\pi <  T < T_{ c}^{*}$  & $\pi$'s & 69 \\
$T_{ c} <  T  < m_{\rm charm}$ \qquad &
  -  $\pi$'s + $  u,{\bar u},d,{\bar d},s,{\bar s}$ + gluons &  247 \\
$m_{ c} <  T < m_\tau$ &  $c,{\bar c}$ & 289 \\
$m_\tau < T < m_{\textrm{bottom}}$ & $\tau {}^{\pm}$ & 303 \\
$m_{ b} < T < m_{ W,Z}$ & $b,{\bar b}$ & 345 \\
$m_{ W,Z} <  T < m_{\textrm{Higgs}}$ & $W^{\pm}, Z$ & 381 \\
$ m_H< T < m_{\textrm{top}}$ & $H^0$ & 385 \\
$m_t< T $ & $t,{\bar t}$  & 427 \\
\hline
\end{tabular}
\end{center}
\vspace{-.2cm}
{\small *$T_{ c}$ 
corresponds to the confinement--deconfinement transition between
 quarks and hadrons. $N(T)$  
 is shown in Fig.~\ref{fig:NT} for $T_{c}  =  150$ and $400$~MeV}.
 \label{tab:NT}
\end{table}

 At higher temperatures, $N(T)$ will be model dependent.
  For example, in the minimal $SU(5)$ model, one needs to add
 to $N(T)$, 24 states for the X and Y gauge bosons, 
another 24 from the adjoint Higgs, and another 6 (in addition
to the 4 already counted in $W^\pm, Z$ and $H$) from the 5
of Higgs.
  Hence for   
$T > M_{ X}  $  in minimal $SU(5)$, $N(T)  =  160.75$. 
 In a supersymmetric model this would at least double, 
with some changes possibly necessary in the table if the
 lightest supersymmetric particle has a mass below  
$m_t$.

\begin{figure}
\centering\includegraphics[width=.7\linewidth]{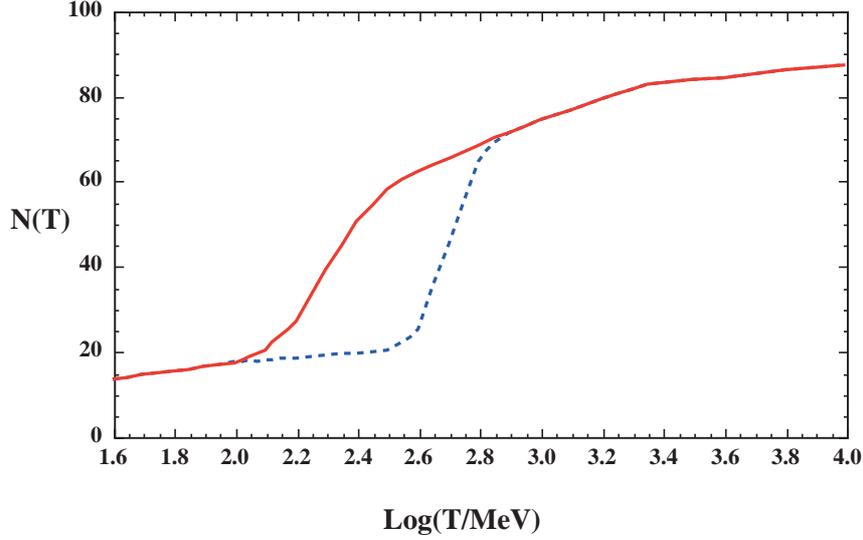}
\caption{The effective numbers of relativistic degrees of freedom
as a function of temperature, assuming a quark--hadron transition temperature of 
150 MeV and 400 MeV (dashed)}
\label{fig:NT}
\end{figure}

The notion of equilibrium also plays an important role in the 
standard Big Bang model. If, for example, the Universe were not
 expanding, then given enough time, each particle state would 
come into equilibrium with every other.  Because of the expansion 
of the Universe, certain rates might be too slow indicating, 
for example, in a scattering process that the two incoming states
 might never find each other to bring about an interaction.  
Depending on their rates, certain interactions may pass in and
 out of thermal equilibrium  during the course of the Universal expansion. 
 Qualitatively, for each particle $i$, we will require that some rate  
$  \Gamma {}_{ i} $  involving that type be larger than the expansion
 rate of the Universe or 
\beq
  \Gamma {}_{ i}  > H ,
\eeq
  in order to be in thermal equilibrium.

A good example for a  process in equilibrium at some stage
 and out of equilibrium at others is that of neutrinos.  
If we consider the standard neutral or charged-current interactions such as 
$e^{ +} + e^{ -  }     \leftrightarrow {}   \nu {}  +  \bar \nu $  
or $e +   \nu {}    \leftrightarrow {}  e  +   \nu {}$  etc.,
 very roughly the rates for these processes will be
 \beq
 \Gamma {} =  n \langle \sigma v \rangle ,
 \eeq
 where  $  \langle \sigma v \rangle $  
is the thermally averaged  weak interaction cross section 
\beq
 \langle \sigma v \rangle { \sim }~\mathcal{O}(10^{ -  2}) T^{ 2} /M_{ W}^4 ,
\eeq
and $n\sim T^3$ is the number density of leptons.
 Hence the rate for these interactions is 
\beq
 \Gamma {}_{\rm wk}   { \sim  }~0(10^{ -  2}) T^{ 5}/M_{ W}^4 .
\eeq 
The expansion rate, on the other hand, is just
\beq
  H  =  \left({8 \pi {}G_{ N}  \rho {} \over 3}\right)^{ 1/2}   
 =  \left({8  \pi {}^{ 3} \over 90}  N(T) \right)^{ 1/2}  T^{ 2}/M_{ P}    
 \sim 1.66 N(T)^{ 1/2}  T^{ 2}/M_{ P} .  
\eeq
The Planck mass $M_{ P} = G_N^{-1/2} =  1.22 \times 10^{19}$ GeV.

Neutrinos will be in equilibrium when  $  \Gamma {}_{\rm wk} >  H $ or
\beq
T > (500 M_{ W}^4)/M_{ P})^{ 1/3}  { \sim }~1 {\rm ~MeV} . 
\eeq
The temperature at which these rates are equal
 is commonly referred to as the decoupling or freeze-out 
temperature and is defined by 
\beq
  \Gamma(T_{ d}) = H(T_{ d}) .
\eeq
 For temperatures $T > T_{ d}$,  
neutrinos will be in equilibrium, while for $T < T_{ d }$ 
 they will not. Basically, in terms of their interactions, 
the expansion rate is just too fast and they never  
{\em `see'}  the rest of the matter in the Universe (nor themselves).
  Their momenta will simply redshift and their effective temperature 
(the shape of their momenta distribution is not changed from that 
of a blackbody) will simply fall with  
$T {~\sim~} 1/R.$

The relation $R T \sim const$ is a direct consequence of the energy conservation 
equation~(\ref{rhod}). Indeed, using $s = dp/dT$, this equation can be rewritten
as 
\beq
\frac{d}{dt} (R^3 s) = 0 ,
\eeq
making it more a statement of conservation of (comoving) entropy than energy (which is not conserved
in comoving coordinates). 

Soon after neutrino decoupling, the $e^{\pm}$  pairs in the thermal background
begin to annihilate (when $T \la m_e$).  
Because the neutrinos are decoupled, 
the energy released heats up the photon background relative 
to the neutrinos. The change in the photon temperature can easily be
computed from entropy conservation. The neutrino entropy 
must be conserved separately from the entropy of interacting particles.
 If we denote $T_>$, the temperature of photons, and $e^{\pm}$
 before annihilation, we also have $T_\nu  = T_>$  as well. 
 The entropy density of the interacting particles at $T = T_>$ is just
\beq
	s_> =  {4 \over 3}  {\rho_> \over T_>}  = 
{4 \over 3} (2 + {7 \over 2}) ( {\pi^2 \over 30} ) T^3_>	,
\eeq
while at $T = T_<$, 
the temperature of the photons just after $e^{\pm}$ annihilation, 
the entropy density is
\beq
	s_< =  {4 \over 3}  {\rho_< \over T_<}  =
 {4 \over 3} (2 ) ( {\pi^2 \over 30} ) T^3_<	,
\eeq
and by conservation of entropy $s_< =  s_>$  and
\beq
	(T_</T_>)^3  = 11/4 .
\eeq
Thus, the photon background is at higher temperature 
than the neutrinos because the $e^{\pm}$
  annihilation energy could not be shared among the neutrinos, and
\beq
	T_\nu = (4/11)^{1/3}   T_\gamma   \simeq 1.9~{\rm K} .
\eeq

For further reading on standard Big Bang cosmology see Refs.~\cite{bbpeeb93,bbborner,bbjap,bbmuk,bbw2,op}.

\subsection{The Cosmic Microwave Background}

There has been a great deal of progress in the last several years
concerning the determination of both $\Omega_m$ and $\Omega_\Lambda$.
Cosmic Microwave Background (CMB) anisotropy experiments have been able
to determine the curvature (i.e., the sum of $\Omega_m$ and
$\Omega_\Lambda$) to better than one per cent, while observations of type Ia
supernovae at high redshift  and baryon acoustic oscillations provide information on (nearly) orthogonal
combinations of the two density parameters.

The CMB is of course deeply rooted in the development and verification of
the Big Bang model and Big Bang Nucleosynthesis (BBN)~\cite{gamo}. 
Indeed, it was the formulation of BBN that
led to the prediction of the microwave background~\cite{ah}.  The argument is rather
simple. BBN requires temperatures greater than 100 keV, which according
to the Standard Model time--temperature relation,
Eq.~(\ref{tT2}), corresponds to timescales less than about 200 s. The typical
cross section for the first link in the nucleosynthetic chain is
\begin{equation} 
\sigma v (p + n \rightarrow D + \gamma) \simeq 5 \times 10^{-20} 
{\rm ~cm}^3/{\rm s} .
\end{equation}
This implies that it was necessary to achieve a density
\begin{equation}
n \sim {1 \over \sigma v t} \sim 10^{17} {\rm ~cm}^{-3} ,
\end{equation}
for nucleosynthesis to begin.
The density in baryons today is known approximately from the density of
visible matter to be ${n_B}_0 \sim 10^{-7}$ cm$^{-3}$ and since
we know that the density $n$ scales as $R^{-3} \sim T^3$, 
the temperature today must be
\begin{equation}
T_0 = ({n_B}_0/n)^{1/3} T_{\rm BBN} \sim 10 {\rm ~K}
\end{equation}
thus linking two of the most important tests of the Big Bang theory.

Of course it was not until many years later that the microwave background radiation
was discovered by Penzias and Wilson~\cite{pw} while perfecting a radio antenna
to track the Echo satellite.  They found a background noise which could
not be eliminated corresponding to a temperature of $3.5 \pm 1$ K. 
One of the most important papers on modern cosmology was published
with the title ``A measurement of excess antenna temperature at 4080-Mc/s''.
This was followed by the seminal paper by Dicke, Peebles, Roll, and Wilkinson~\cite{dprw}
putting this observation in a cosmological context.
Subsequently, there have been many observations of the CMB
culminating in the COBE observation~\cite{cobe}
which determined the temperature to an unprecedented level, 
set aside any lingering doubts about the true black body nature of the CMB,
and discovered the intrinsic anisotropies in the background.
 
An enormous amount of cosmological information is encoded in the
angular expansion of the CMB temperature
\begin{equation}
T(\theta,\phi) = \sum_{\ell m} a_{\ell m} Y_{\ell m}(\theta, \phi) .
\end{equation}
The monopole term characterizes the mean background temperature
of $T_\gamma = 2.725 \pm 0.001$ K as determined by COBE~\cite{cobem},
whereas the dipole term can be associated with the Doppler shift
produced by our peculiar motion with respect to the CMB.
In contrast, the higher order multipoles are directly related to
energy density perturbations in the early Universe. When compared
with theoretical models, the higher order anisotropies can be used
to constrain several key cosmological parameters.
In the context of simple adiabatic cold dark matter (CDM) models,
there are nine of these: the cold dark matter density, $\Omega_m h^2$;
the baryon density, $\Omega_B h^2$; the curvature --- characterized by $\Omega_{\rm total}$;
the hubble parameter, $h$; the optical depth, $\tau$; the spectral indices of scalar and 
tensor perturbations, $n_s$ and $n_t$; the ratio of tensor to scalar perturbations, $r$;
and the overall amplitude of fluctuations, $Q$.

\begin{figure}
\centering\includegraphics[width=.90\linewidth]{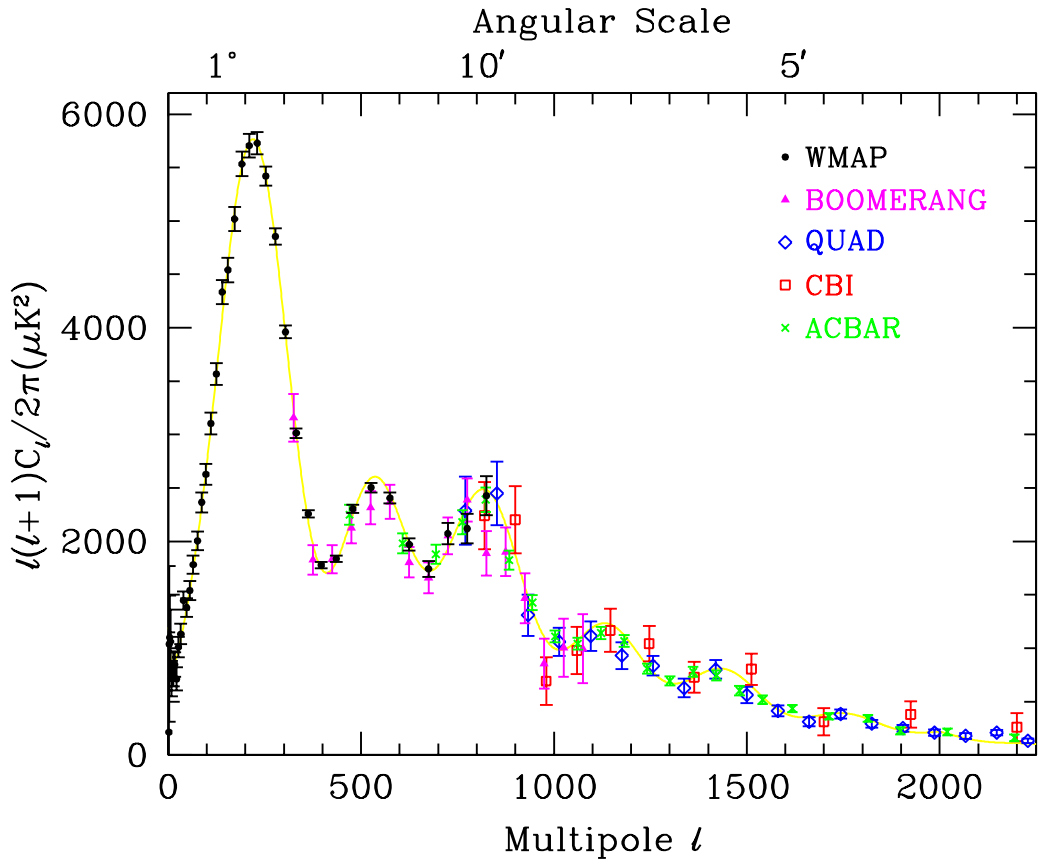}
\caption{The power in the microwave background anisotropy spectrum
as measured by WMAP~\protect\cite{wmap}, Boomerang~\protect{\cite{boom}}, 
QUaD~\protect{\cite{quad}}, CBI~\protect\cite{cbi}, and ACBAR~\protect\cite{acbar}.
Figure courtesy of D. Scott~\cite{scsm}. }
\label{freq1}
\end{figure}

Microwave background anisotropy measurements have made tremendous
advances in the last few years. The power spectrum~\cite{max,boom,dasi1,cbi,vsa,arch,wmap0,acbar,quad}
has been measured relatively accurately out to multipole moments
corresponding to $\ell \sim 2000$. A compilation of recent data is shown in Fig. 
\ref{freq1}~\cite{scsm}, where the power at each ${\ell}$ is given by $(2{\ell} + 1) C_\ell/(4 \pi)$, 
and $C_\ell = \langle |a_{\ell m}|^2 \rangle$.

As indicated above, the details of this spectrum enable
one to make accurate predictions of a large number of fundamental
cosmological parameters. The results of the WMAP data 
(with other information concerning the power spectrum) is shown in Table~\ref{tab:wmap}.
For details see Refs.~\cite{wmap,scsm,ll}.

\begin{table}[h]
\centering
\caption{WMAP determinations of cosmological parameters}
\begin{tabular}{l|cc}
& {\bf WMAP alone} & {\bf WMAP + BAO + SN} \\
\hline
$\Omega_{{\rm m}} h^2$ & $0.133 \pm 0.006$ & $0.136 \pm 0.004$\\
$\Omega_{{\rm B}} h^2$ & $0.0227 \pm 0.0006$ & $0.0227 \pm 0.0006$\\
$h$ & $0.72 \pm 0.03$ & $0.705 \pm 0.013$\\
$n_s$ & $0.963^{+ 0.014}_{-0.015}$ & $0.960 \pm 0.013$\\
$\tau$ & $0.087 \pm 0.017$ & $0.084 \pm 0.016$\\
$\Omega_\Lambda$ & $0.74 \pm 0.03$ & $ 0.726 \pm 0.015$ \\
\end{tabular}
\label{tab:wmap}
\end{table}

\begin{figure}
\centering\includegraphics[width=.7\linewidth]{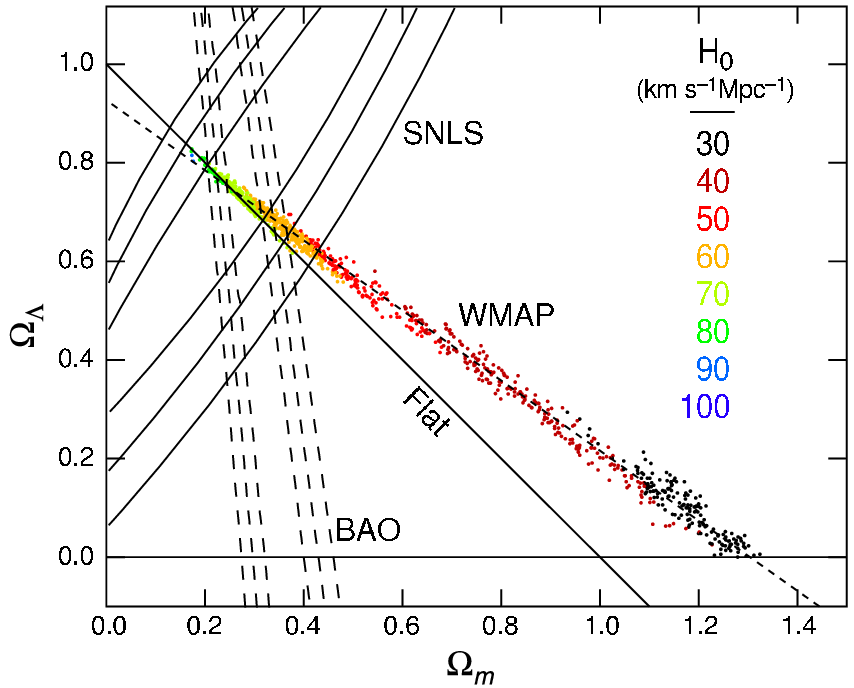}
\caption{\footnotesize Two-dimensional confidence regions
in the ($\Omega_m,\Omega_{\Lambda}$) plane.
The coloured Monte Carlo
points derive from WMAP~\cite{wmap0} and show that the CMB alone
requires a flat Universe $\Omega_{\Lambda}+\Omega_{m}\simeq 1$ if the Hubble constant is 
not too high. The SNe Ia results~\cite{sn2} very nearly 
constrain the orthogonal combination $\Omega_{\Lambda}-\Omega_{m}$.
Also shown is the region obtained from baryon acoustic oscillations~\cite{kow}.
The intersection of these constraints is the most direct 
(but far from the only) piece of evidence favouring a 
flat model with $\Omega_{m}\simeq 0.25$. }
\label{lm}
\end{figure}

Of particular interest to us here is the CMB determination of the total
density $\Omega_{\rm tot}$ as well as the matter density $\Omega_m$. 
There is strong evidence that the Universe is flat or
very close to it. As noted earlier, the best determination of $\Omega_{\rm total}$ is $1.006 \pm 0.006$. 
Furthermore, the matter density is significantly larger than the baryon
density implying the existence of cold dark matter. Also,  the baryon density,
as we will see below, is consistent with the BBN production of D/H and its
abundance in quasar absorption systems. 
The apparent discrepancy between the CMB value
of $\Omega_{\rm tot}$ and $\Omega_m$, though not conclusive on its own,
is a sign that a contribution from the vacuum energy density or
cosmological constant, is also required.  The preferred region in the
$\Omega_m - \Omega_\Lambda$ plane is shown in Fig.~\ref{lm}.

The presence or absence of a cosmological constant is a long standing
problem in cosmology.   We know that 
the cosmological term is at most a factor of a few times larger than the
current mass density.  Thus from Eq.~(\ref{H}), we see that the
dimensionless combination $G_N \Lambda \la 10^{-121}$. 
Also shown in Fig.~\ref{lm}, are 
the results from SN~Ia~\cite{sn1,sn2} and baryon acoustic oscillations~\cite{bao}.
Taken together, 
we are led to a seemingly conclusive
picture.  The Universe is nearly flat with $\Omega_{\rm tot} \simeq 1$.
However, the density in matter makes up only 23\% of this total, with
the remainder in a cosmological constant or some other form of dark
energy.

\section{Lecture 2: Inflation and Baryogenesis}

Despite the successes of the standard Big Bang model, there are a
number of unanswered questions that appear difficult to explain without
imposing what may be called unnatural initial conditions.  
The resolution of these problems may lie in a unified theory of gauge 
interactions or possibly in a theory which includes gravity.
For example, prior to the advent of grand unified theories (GUTs),
the baryon-to-photon ratio, could have been viewed as 
being embarrassingly small.
Although we still do not know the
 precise mechanism for generating the baryon asymmetry, many
quite acceptable models
are available. 
In a similar fashion, it is hoped that a field theoretic
description of inflation may resolve the problems outlined below. 

\subsection{Cosmological problems}
	\subsubsection{The curvature problem}
As noted above, the determined value of $\Omega$ in Eq.~(\ref{range}) is 
curious since at the present
time we do not know even the sign of the curvature term in the Friedmann
equation~(\ref{H}), i.e., we do not know if the Universe is open, closed or
spatially flat.  

	The curvature problem (or flatness problem or age problem) 
can manifest itself in
several ways. For a radiation
dominated gas, the entropy density $s \sim T^3$  and $R \sim T^{-1}$. 
 Thus assuming an adiabatically expanding
Universe, the quantity ${\hat k} = k/R^2T^2$ is a dimensionless constant.
  If we now
apply the limit in Eq.~(\ref{range}) to Eq.~(\ref{o-1}) (with $\Lambda = 0$) we find
\beq
{\hat	k} = {k \over R^2 T^2}  = {(\Omega_0  - 1){H_0}^2 \over {T_0}^2} 
  \la 10^{-60} .
  \label{khat}
\eeq
This limit on $\hat k$ represents an initial condition on the cosmological model.
The problem then becomes what physical processes, if any, in the early Universe
produced a value of $\hat{k}$ so extraordinarily close to zero 
(or $\Omega$ close to one).
	A more natural initial condition might have 
been $\hat{k} \sim 0(1)$.  In this
case the Universe would have become curvature 
dominated at $T \sim 10^{-1} M_{\rm P}$.
  For
$k = +1$, this would signify the onset of recollapse.
As already noted earlier, one would naturally 
expect the effects of curvature
(seen in Fig.~\ref{fig:myfig} by the separation of the 
three curves) to manifest themselves at times on the order
of the Planck time as gravity  should provide the only 
dimensionful scale in this era. 
If we view the evolution of $\Omega$ in Fig.~\ref{fig:myfig}
as a function of time, then it would appear that the time
$t_0 = 13.7$ Gyr = $8 \times 10^{60} {M_{\rm P}}^{-1}$ ($\sim$ the current
 age of the Universe) appears at the far left of the x-axis, i.e., before the
curves separate. Why then has the Universe lasted so long before revealing
the true sign of $k$?

\subsubsection{The horizon problem}

	Because of the cosmological principle, all physical length scales grow
as the scale factor $R(t) \sim t^{2/3\gamma}$, with $\gamma$ defined by
$p = (\gamma - 1) \rho$. 
 However, as we have seen, there is 
a particle horizon $d_H(t) \sim t$ as defined in Eq.~(\ref{dh}). 
 For $\gamma > {2 \over 3}$,
 scales originating outside of the horizon will eventually
become part of our observable Universe.  Hence we would expect to see
anisotropies on large scales~\cite{hor}.

	In particular, let us consider the microwave background today.  The
photons we observe have been decoupled 
since recombination at $T_d \sim 3000$~K.
At that time, the horizon volume was simply $V_d \propto {t_d}^3$,
 where $t_d$  is the age of
the Universe at $T = T_d$.  
Then $t_d  = t_0 (T_0/T_d)^{3/2}    \sim 2 \times 10^5$ yrs, where
$T_0 =  2.725$~K~\cite{cobem} is the present temperature of the
microwave background. 
 Our present horizon
volume $V_0 \propto {t_0}^3$  can be scaled back to
 $t_d$ (corresponding to that part of the
Universe which expanded to our present visible 
Universe) $V_0(t_d) \propto V_0
(T_0/T_d)^3$.  We can now compare $V_0(t_d)$ and $V_d$.  The ratio
\beq
	{V_0(t_d) \over V_d}
 \propto {V_0 {T_0}^3 \over V_d {T_d}^3}
  \propto {{t_0}^3 {T_0}^3 \over {t_d}^3 {T_d}^3}  \sim 5 \times 10^4 	
\eeq
corresponds to the number of horizon volumes or casually distinct regions at
decoupling which are encompassed in our present visible horizon.

	In this context, it is astonishing that the microwave
 background appears highly
isotropic on large scales with
 $\Delta T/T = 1.1 \pm 0.2 \times 10^{-5}$ 
at angular separations of $10^{\circ}$~\cite{cobe}.  The horizon
problem, therefore, is the lack of an explanation as to why nearly $10^5$
causally disconnected regions at recombination all had the same temperature
to within one part in $10^{-5}$.

\subsubsection{Density perturbations}

	Although it appears that the Universe is extremely isotropic and
homogeneous on very large scales 
(in fact the Standard Model assumes complete isotropy and
homogeneity) it is very inhomogeneous on small scales.  In other words,
there are planets, stars, galaxies, clusters, etc. 
 On these small scales there
are large density perturbations namely $\delta \rho/\rho \gg 1$.
At the same time, we know from the isotropy of the microwave background
that on the largest scales, $\delta \rho /\rho \sim 3 \Delta T/ T
\sim O(10^{-5})$~\cite{cobe}, and these perturbations must have grown to
$\delta \rho/\rho \sim 1$ on smaller scales.

	In an expanding Universe, density perturbations evolve with time~\cite{bbpeeb93}.
 The evolution of  the Fourier transformed quantity
 ${\delta \rho \over \rho}(k,t)$ depends on
the relative size of the wavelength $\lambda \sim k^{-1}$ 
 and the horizon scale $H^{-1}$.  For
$k \ll H$, (always true at sufficiently early times)
$\delta \rho/\rho \propto t$ while for $k \gg H$,
$\delta \rho/\rho$ is $\simeq$ constant (or grows moderately
as $\ln t$)
 assuming a radiation dominated Universe.
In a matter dominated Universe, on scales larger than the Jean's 
length scale (determined by $k_J = 4\pi G_N \rho_{matter}/{v_s}^2$, $v_s$ =
sound speed) perturbations grow with the scale factor $R$.
  Because of the
growth in $\delta \rho/\rho$, the microwave background 
limits force $\delta \rho/\rho$ to be extremely
small at early times.

	Consider a perturbation with wavelength on the order of a galactic
scale.  Between the Planck time and recombination, such a perturbation would
have grown by a factor of $O(10^{57})$ and the anisotropy limit 
of $\delta \rho/\rho \la 10^{-5}$
implies that $\delta \rho/\rho < 10^{-61}$ on the scale
 of a galaxy at the Planck time.  One
should compare this value with that predicted from purely random (or
Poisson) fluctuations of $\delta \rho/\rho \sim 10^{-40}$
   (assuming $10^{80}$  particles (photons) in
a galaxy)~\cite{bg}.  The extent of this limit 
is of course related to the fact
that the present age of the Universe is so great.

	An additional problem is related to the formation time of the
perturbations.  A perturbation with a wavelength large enough to correspond
to a galaxy today must have formed with wavelength modes much greater than
the horizon size if the perturbations are primordial, as is generally
assumed.  This is due to the fact that the wavelengths red-shift as $\lambda
\sim R \sim t^{1/2}$ while the horizon size grows linearly. 
It would appear that a mechanism for generating 
perturbations with acausal wavelengths is required.

\subsubsection{The magnetic monopole problem}

	In addition to the much desired baryon asymmetry produced by grand
unified theories, a less favourable aspect is also present; GUTs predict the
existence of magnetic monopoles.  Monopoles will be produced~\cite{thp}
whenever any simple group [such as $SU(5)$] is broken down to a gauge group
which contains a $U(1)$ factor [such as $SU(3) \times SU(2) \times U(1)$]. 
 The mass of
such a monopole would be
\beq
	M_m \sim M_{GUT}/\alpha_{GUT}  \sim 10^{16}  {\rm ~GeV}.		
\eeq
The basic reason monopoles are produced is that in the breaking of SU(5), the
Higgs adjoint needed to break SU(5) 
cannot align itself over all space~\cite{kib}. 
 On scales larger than the
horizon, for example, there is no reason to expect the direction of the
Higgs field to be aligned.  Because of this randomness, topological knots
are expected to occur and these are the magnetic monopoles.  We can then
estimate that the minimum number of monopoles produced~\cite{mon}
 would be roughly
one per horizon volume or causally connected region at the time of the $SU(5)$
phase transition $t_c$,
\beq
	n_m  \sim (2t_c)^{-3}  	,	
\eeq
resulting in a monopole-to-photon ratio expressed in terms of the transition
temperature of
\beq
	{n_m \over n_\gamma} \sim \left({10T_c \over M_{\rm P}}\right)^3 	.
\label{mop}	
\eeq
	The overall mass density of the Universe can be used to place a
constraint on the density of monopoles. 
For $M_m  \sim 10^{16}$ GeV and $\Omega_m {h_0}^2  \la 1$ we have that
\beq
	{n_m \over n_\gamma}  \la 0(10^{-25}) .
\eeq
The predicted density, however, from Eq.~(\ref{mop}) for $T_c \sim M_{GUT}$ is
\beq
	{n_m \over n_\gamma}  \sim 10^{-9} .
\eeq
Hence we see that standard GUTs and cosmology have a monopole problem.

\subsection{Inflation}
All of the problems discussed above can be neatly resolved if the 
Universe underwent a period of cosmological inflation~\cite{guth,infl}.
That is, if the Universe at some stage becomes dominated by
the vacuum as could be the case
during a phase transition, our assumptions of an
adiabatically expanding Universe may not be valid.  
Indeed, we expect several cosmological phase transitions to 
have occurred including the breakdown of a Grand Unified symmetry such
as SU(5) $\to$ SU(3)$_c \times$ SU(2)$_L \times$ U(1)$_Y$ or the 
electroweak transition SU(2)$_L \times$ U(1)$_Y \to$ U(1)$_{em}$, or possibly some
other non-gauged transition. 

During a phase transition, the motion of a scalar field will be
described by a scalar potential.
If the solution to the equations of motion for the
scalar field leads to a slowly evolving scalar field
(this will depend on the details of the 
potential), the Universe may become dominated
by the vacuum energy density associated with 
the potential near the initial field value, say $\phi \approx 0$.
The energy density of the symmetric vacuum $V(0)$ acts as a cosmological 
constant with
\beq
	\Lambda = 8\pi V(0)/ {M_{\rm P}}^2	.
\eeq
During this period of slow evolution, the energy 
density due to radiation or matter will fall below the vacuum 
energy density, $\rho \ll V(0)$.  When this happens, the expansion 
rate will be dominated by the constant $V(0)$ and from Eq.~(\ref{H})
we find the exponentially expanding solution given in Eq.~(\ref{Vdom}).
When the field evolves towards the global minimum 
it will begin to oscillate
about the minimum, energy will be released
during its decay and a hot thermal Universe will be restored.
 If released 
fast enough, it will produce radiation at a temperature 
${T_R}^4 \la V(0)$.  In this reheating process, entropy has been created and
	$(RT)_f  > (RT)_i $.	
Thus we see that during a phase
 transition, the relation $RT \sim$ constant need not hold true and 
our dimensionless constant $\hat{k}$ may actually not have been constant.

If during the phase transition  the value of
 $RT$ changed by a factor of ${\mathcal O}(10^{30})$, the cosmological
 problems discussed above would be solved.  The isotropy would in a sense be
 generated by the immense expansion; one small causal region
 would get blown up and our entire visible Universe would 
have been at one time in thermal contact.  In addition, 
the parameter $\hat{k}$ could have started out ${\mathcal O}(1)$ and have
 been driven small by the expansion. The wavelengths of density perturbations
would have been stretched by the expansion $\lambda \sim R$
making it appear that $\lambda \gg H^{-1}$ or that the perturbations
have left the horizon.  Rather, it is the size of the causally
connected region that is no longer simply $H^{-1}$. However, not only does
inflation offer an explanation for large scale perturbations, it also 
offers a source for the perturbations themselves~\cite{pert}.
Monopoles would also be diluted away.

The cosmological problems could be solved if
\beq
	H\tau > 65
\eeq
where $\tau$ is the duration of the phase 
transition. In a successful theory, density perturbations are
produced and do not exceed the limits imposed by
the microwave background anisotropy,  the vacuum energy density was converted
 to radiation so that the reheated temperature is sufficiently
high, and baryogenesis is realized.

In the original (old) inflationary scenario, the phase transition
 determined by a potential with a large barrier 
separating the false and true vacua proceeds via the formation of bubbles~\cite{col}.
  The Universe reheats with the release of entropy which must 
occur through bubble collisions and the transition is completed 
when the bubbles fill up all of space.  It is known~\cite{gw},
 however, that the requirement for a long timescale $\tau$
 is not compatible with the completion of the phase transition.
  The Universe as a whole remains trapped in the exponentially 
expanding phase containing only a few isolated bubbles 
of the broken $SU(3) \times SU(2) \times U(1)$ phase. 

The well-known solution to the dilemma of old inflation is called the
 new inflationary scenario~\cite{new}.
  New inflation (it was at the time) was originally based on symmetry breaking 
using a flat potential of the Coleman--Weinberg form Ref.~\cite{loop}.
Instead of proceeding by tunnelling and the formation of bubbles,
the transition takes place more or less uniformly on large scales.
The details of the inflationary transition are determined from the equations of motion.

	A Lagrangian for a scalar field which includes a scalar potential
may be incorporated into the total action including gravity
\beq
I = \int d^4x \sqrt{g} \left(\frac{R}{2 \kappa^2}
 - \frac{1}{2} \partial_{\mu} \phi
\partial^{\mu} \phi - V(\phi)\right) .
\eeq
The equation of
motion for a scalar field $\phi$ can be derived from the energy-momentum tensor
\beq
	T_{\mu\nu}  
	    = \partial_{\mu} \phi \partial_{\nu} \phi 
- {1 \over 2} g_{\mu\nu}  \partial_{\rho} \phi
\partial^{\rho} \phi - g_{\mu\nu}V(\phi)	.
\eeq
By associating $\rho = T_{00}$   and $p = R^{-2}(t)T_{ii}$   we have
\begin{eqnarray}
	\rho =  {1 \over 2} {\dot{\phi}}^2  + {1 \over 2} R^{-2}(t)
(\nabla\phi)^2  + V(\phi),		\\
	p =  {1 \over 2} {\dot{\phi}}^2  - {1 \over 6} R^{-2}(t)
(\nabla\phi)^2  - V(\phi)	,
\label{prho}
\end{eqnarray}
and from Eq.~(\ref{rhod}) we can write the equation of motion (by considering a
homogeneous region, we can ignore the gradient terms)
\beq
	\ddot{\phi} + 3H\dot{\phi} = -\partial V/\partial \phi .
\label{eom}
\eeq

Consider now the approximation $\partial V/\partial \phi \sim
(\partial^2 V/\partial \phi^2) \phi$; the equation of motion becomes~\cite{lin1}
\beq
	\ddot{\phi} + 3H\dot{\phi} + m^2(\phi)\phi = 0	,	
\eeq
with $m^2(\phi) = \partial^2 V/\partial \phi^2  < 0$. 
 The solution when $|m^2| \gg H^2$  grows exponentially
as $\phi \sim e^{|m|t}$ while for $|m^2| \ll H^2$  
the scalar field grows as $\phi \sim e^{|m|^2t/3H}$.
In the latter case the field moves very slowly during a time period 
\beq
\tau \sim 3H/|m|^2 .
\label{efold}
\eeq
This approximation is known as the slow-rollover approximation.

If the scalar mass is tuned somewhat, $m_\phi \sim 10^9$ GeV,
a significant amount of inflation is possible.
From Eq.~(\ref{efold}) one sees that
\beq
H\tau \sim {H^2 \over |m|^2} \sim {v^4 \over {M_{\rm P}}^2 |m|^2} \sim 10^4 ,
\eeq
for $v \sim 10^{15}$ GeV.
Reheating no longer occurs via the collisions of bubbles,
but by the decay of scalar field oscillations. As the scalar field
settles to its minimum, the solution to the equations of motion
look like
\beq
\phi(t) \sim {v \over mt} \sin mt
\eeq
and the reheat temperature is
\beq 
T_R \sim \left(\Gamma_D M_{\rm P}\right)^{1/2} \qquad \Gamma_D < H_I ,
\eeq
where $\Gamma_D$ is the scalar field decay rate and $H_I$ is the 
value of $H$ during inflation.
 
 In addition to producing $\Omega = 1$, which is clearly seen from 
Eq.~(\ref{khat}) as $\hat{k} \rightarrow 0$, new inflation is capable of
 producing scale invariant density perturbations~\cite{pert} of the type
 preferred for galaxy formation models.  However, the original~\cite{new} 
new inflationary models based on a Coleman--Weinberg~\cite{loop} type of $SU(5)$
 breaking produced density fluctuations with magnitude $\delta\rho/\rho
 \sim O(10^2)$ rather 
than $\delta\rho/\rho \sim 10^{-5}$
 as needed to remain consistent with microwave background
 anisotropies. Other more technical problems~\cite{lin2}
 concerning slow rollover and
the effects of quantum fluctuations also pass doom on this original model.

General models of inflation can be described by a few so-called
slow roll parameters $\epsilon$ and $\eta$.  These are given by
\beq 
\epsilon = \frac{3}{2} \left( \frac{p}{\rho} + 1 \right) = \frac{4 \pi}{M_{\rm P}^2} \left( \frac{\dot \phi}{H} \right)^2 
\simeq  \frac{M_{\rm P}^2}{16 \pi}  \left( \frac{V^\prime(\phi)}{V(\phi)} \right)^2
\eeq
and 
\beq
\eta = - \frac{\ddot \phi}{H {\dot \phi} } \simeq  \left( \frac{V^{\prime\prime}(\phi)}{V(\phi)} \right) .
\eeq
For sufficient inflation both of these parameters must be small. 
The amount of inflation (given by the total number of e-foldings, $N$) is given by
$\epsilon$,
\beq
N = - \int H dt = \int \frac{H}{\dot \phi} d \phi = \frac{2 \sqrt{\pi}}{M_{\rm P}}\int \frac{d\phi}{\sqrt{\epsilon}} .
\eeq

As noted above, inflation leads to a nearly scale-free spectrum of density fluctuations,
$\Delta(k) \propto \delta\rho/\rho$ with a power spectrum of the form
$\Delta^2 (k) \propto k^{n-1}$. The spectral index is determined by the inflationary potential
\beq
n \simeq 1 - 6 \epsilon + 2\eta .
\eeq
In addition to the scalar perturbations, tensor perturbations (gravity waves)
will also be produced during inflation with spectral index $n_g \simeq -2 \epsilon$.
The ratio of the amplitudes of the tensor to scalar perturbations is an important
observable also given in terms of the slow roll parameter, $\epsilon$, $r \simeq 16 \epsilon$.

A generic model of inflation can be described by a potential
of the form
\begin{equation}
 V( \phi ) = {{\mu}^4} P( \phi ) ,
\label{a}
\end{equation}
where $\phi$ is the scalar field driving inflation, the inflaton,  
$\mu$ is an as yet unspecified mass parameter, and $P(\phi)$ is a  
function of $\phi$ which possesses the features necessary for  
inflation, but contains no small parameters. 
That is,  $P(\phi)$ takes the form
\beq
P(\eta) = P(0) + m^2 \eta^2 + \lambda_3 \eta^3 + \lambda_4 \eta^4   +...   ,
\eeq
where all of the couplings in $P$ are $O(1)$ and $...$ refers to 
possible non-renormalizable terms. Most of the useful  
inflationary potentials can be put into the form of Eq.~(\ref{a}).

The requirements for successful inflation boil down to: 1) enough inflation; and
2) density perturbations of the right magnitude. The latter
reduces approximately to
\beq
{\delta \rho \over \rho} \sim O(100) {\mu^2 \over {M_{\rm P}}^2} .
\label{drho}
\eeq
  For large scale fluctuations of the type measured by COBE~\cite{cobe},
we can use Eq.~(\ref{drho}) to fix the inflationary scale $\mu$
of the inflaton potential~\cite{cdo}:
\begin{equation}
{\frac{\mu^2}{M_{\rm P}^2} = {\rm few} \times{10^{-8}}} .
\label{cobemu}
\end{equation}

       Fixing $({\mu^2}/{M_{\rm P}^2})$ has immediate general consequences  
for inflation~\cite{eeno}. For example, the Hubble parameter during inflation,  
${{H^2} \simeq (8\pi/3)({\mu^4}/{M_{\rm P}^2})}$ so that $H \sim  
10^{-7}M_{\rm P}$. The duration of inflation is $\tau \simeq  
{M_{\rm P}^3}/{\mu^4}$, and the number of e-foldings of expansion is $H\tau  
\sim 8\pi({M_{\rm P}^2}/{\mu^2}) \sim 10^{9}$. If the inflaton decay rate  
goes as $\Gamma \sim {m_{\eta}^3}/{M_{\rm P}^2} \sim {\mu^6}/{M_{\rm P}^5}$, the  
Universe recovers at a temperature $T_R \sim (\Gamma{M_{\rm P}})^{1/2} \sim  
{\mu^3}/{M_{\rm P}^2} \sim 10^{-11} {M_{\rm P}} \sim 10^8 {\rm  ~GeV}$. 

Two commonly studied potentials are associated with a form of inflation known as chaotic
inflation~\cite{lin3}. In these models it is assumed that as part of
an initially chaotic state $\phi > M_{\rm P}$  with $V(\phi) \sim M_{\rm P}^4$. 
Once these
assumptions are made, chaotic models of inflation are by far the simplest.
	Typical models for chaotic inflation in terms of a single scalar field
are described by the following scalar potentials~\cite{lin3,lin4}
\beq
	V =  {1 \over 4} \lambda \phi^4
\eeq	
or
\beq
	V =  {1 \over 2} m^2  \phi^2 	.
\eeq
That's all!  Nothing more complicated is necessary.   It is
assumed that at the Planck time, all fields $\phi(x)$
 satisfy $V(\phi) < {M_{\rm P}}^4$  and
$(\partial_\mu \phi)^2  < {M_{\rm P}}^4$.  
It is also assumed that there exist domains sufficiently
large $l > H^{-1}$   with $\phi$ homogeneous and $\phi > M_{\rm P}$.

For suitably large initial values of the scalar field $\phi$, the Universe
expands quasiexponentially. Sufficient inflation requires only that
$\phi_o > $ few $\times M_{\rm P}$.  
	Although this is not a strong constraint on
chaotic models, a stronger constraint is derivable from the consideration of
density perturbations.  
 One finds that $\delta\rho/\rho \la 10^{-5}$   for $\lambda < 10^{-14}$
or $m < 10^{-6}M_{\rm P}$. 
The slow roll parameters are easily determined for these two models and are given in Table~\ref{slow}.

\begin{table}
\centering
\caption{Slow roll parameters for chaotic inflationary models}
\begin{tabular}{lcc}
& \boldmath$V(\phi) = m^2 \phi^2$  & \boldmath$V(\phi) = \lambda \phi^4$\\
\hline
$\epsilon$ & 1/120 & 1/60\\
$\eta$ & 1/120 & 1/40\\
$n$ & 0.97 & 0.95 \\
$r$ & 0.13 & 0.27\\
\end{tabular}
\label{slow}
\end{table}

Finally, CMB measurements can be used to test inflationary models
by determining or limiting the slow roll parameters.
For example, WMAP~\cite{wmap} is able to set a constraint
in the $r,n$ plane as shown.  For example, at $n = 0.95$, the 68\% (95\%) CL upper limit
on $r$ is $<$ 0.07 (0.17), while at $n = 0.97$, $r < $ 0.18 (0.28).
As one can see, while the $m^2 \phi^2$ model is well within the constraints,
the $\lambda \phi^4$ model is not.

\subsection{Baryogenesis}

It appears
that there is apparently very little antimatter 
in the Universe.  
To date, the only antimatter
observed is the result of a high-energy collision, either in an 
accelerator or in a cosmic-ray collision in the atmosphere. There has been no
sign to date of any primary antimatter, such as an 
anti-helium nucleus ${\bar \alpha}$
found in cosmic-rays.
In addition, 
the number of photons greatly exceeds
 the number of baryons. Indeed, the value of $\Omega_B h^2$ as determined by
 WMAP~\cite{wmap} and listed in Table~\ref{tab:wmap} corresponds to a baryon-to-photon
 ratio of 
 \beq
 \eta = \frac{n_B -n_{\bar B}}{n_\gamma} \simeq \frac{n_B}{n_\gamma} \simeq 274 \Omega _B h^2 = (6.23 \pm 0.17) \times 10^{-10} .
 \label{etas}
 \eeq
 In the Standard Model, the entropy density today is related to
$n_\gamma {}$  by
\beq
s  =  \frac{2 \pi^4}{90 \zeta(3)} ( 2+ \frac{21}{4} \frac{4}{11} )n _\gamma =  7.04 n_\gamma ,
\eeq  
so that Eq.~(\ref{etas}) implies $n_{ B}/s  \sim 8.8 \times 10^{-11}$. 
In the absence of baryon number violation or entropy production 
this ratio is conserved, however, and hence represents a potentially undesirable 
initial condition.

Let us for the moment assume that in fact  $\eta  = $  0.  We can 
compute the final number density of nucleons 
left over after annihilations of baryons and 
antibaryons  have frozen out.  At very high 
temperatures (neglecting a quark--hadron
 transition) $T  
>$  1 GeV, nucleons were in thermal equilibrium with 
the photon background and $n 
_{B} = n_{\bar B} = (3/2)n_\gamma$  (a factor of 2 accounts
 for neutrons and protons and the factor 3/4 
 for the difference between Fermi and Bose statistics). 
 As the temperature fell below $m_N$
  annihilations kept the nucleon density at its equilibrium value 
$(n_B/n_\gamma) = (\pi^{1/2} (m_{N}/T)^{3/2} / 2^{3/2} \zeta(3)){\rm exp}(-m_{N}/T)$ 
 until the annihilation rate  
$\Gamma_A \simeq n_B m_\pi^{-2}$ 
 fell below the expansion rate. This occurred at $T  
\simeq$  20~MeV.  However, at this time the nucleon 
number density had already dropped to
\beq
n_B/n_\gamma = n_{\bar B}/n_\gamma \simeq 10^{-18} ,
\eeq
 which is eight orders of magnitude too small~\cite{Gary} aside from 
the problem of having to separate the baryons from the antibaryons.
 If any separation did occur at higher temperatures 
(so that annihilations were 
as yet incomplete) the maximum distance scale on which separation could occur 
is the causal scale related to the age of the Universe at that time.  At $T  
=$  20 MeV, the age of the Universe was only $t  =  2 \times 10^{-3}$~s.  
At that time, a causal region (with distance scale defined by $2ct$) 
could only have contained 
$10^{-5} M_\odot$  which is very far from the galactic mass scales, $10^{12} M_\odot$,
we are asking for separations to occur.

\subsubsection{The out-of-equilibrium decay scenario}

The production of a net baryon asymmetry requires baryon number violating
interactions, C and CP violation and a departure 
from thermal equilibrium~\cite{sak}.
The first two of these ingredients are contained in GUTs, 
the third can be realized in an expanding Universe
 where, as we have seen, it is not uncommon that interactions 
come in and out of equilibrium.  
In SU(5), the fact that quarks and leptons are in the same multiplets allows
 for baryon non-conserving interactions such as 
$e^{-} + d  \leftrightarrow {\bar u} + {\bar u}$,  etc., 
or decays of the supermassive
 gauge bosons $X$ and $Y$ such as 
$ X  \rightarrow e^{-} + d, {\bar u} + {\bar u}$. 
 Although today these interactions 
are very ineffective because of the very large masses of the $X$ 
and $Y$ bosons, in the early Universe when   
$T \sim M_{ X} \sim 10^{15}$  GeV these types of interactions 
should have been very important.
 C and CP violation is very model dependent.  In the minimal SU(5) 
model, as we will see,
the magnitude of C and CP violation is too small to yield a useful value of  
$\eta$.  The C and CP violation in general  comes 
from the interference between
 tree level and first loop corrections.

The departure from equilibrium is very common in the 
early Universe when interaction 
rates cannot keep up with the expansion rate.  In fact, 
the simplest (and most useful) 
scenario for baryon production makes use of the fact that a 
single decay rate goes out of equilibrium.  It is commonly referred to 
 as the out-of-equilibrium decay scenario~\cite{ww}.  The basic idea is that the gauge bosons
 $X$ and $Y$ (or Higgs bosons)
 may have a lifetime long enough to insure that the 
inverse decays have already 
ceased so that the baryon number is produced by their free decays.
 
More specifically, let us call $X$, either the gauge 
boson or Higgs boson which produces 
the baryon asymmetry through decays.  Let  
$\alpha$  be its coupling to fermions.  For $X$ a gauge boson,  $\alpha$ 
will be the GUT fine structure constant, while for $X$ a Higgs boson,  
$(4{\pi \alpha })^{ 1/2}$  will be the Yukawa coupling to fermions. 
 The decay rate for $X$ will be  
\beq
 \Gamma_{ D}  \simeq   \alpha M_{X} .
\eeq
  However, decays can only begin occurring when the age 
of the Universe is longer
 than the $X$ lifetime   
$\Gamma_D^{-1}$,  i.e., when  $\Gamma_{ D} >  H$  
\beq
  \alpha M_{ X}  \ga  N(T)^{ 1/2} T^2/M_{ P} ,
\eeq
 or at a temperature 
\beq
 T^{ 2}  \la  \alpha M_{ X}M_{ P}N(T)^{ -1/2}.
\eeq
Scatterings, on the other hand, proceed at a rate  
$\Gamma_{ S}  \sim \alpha^2 T^3/M_X^2$ 
 and hence are not effective at lower temperatures.  To be in equilibrium,
decays must have been effective as $T$ fell below 
$M_{ X}$  in order to track the equilibrium 
density of $X$'s (and  ${\bar X}$'s). 
Therefore, the out-of-equilibrium condition  is 
that at $T = M_{ X},   \Gamma {}_{ D} < H$  
or
 \beq
M_{ X} \ga  \alpha M_{ P} (N(M_{ X}))^{ -1/2}  
\sim 10^{18} \alpha {\rm ~GeV} .
\label{mxmin}
\eeq
 In this case, we would expect a maximal net baryon asymmetry to be produced.

To see the role of C and CP violation, consider the 
two channels for the decay
of an $X$ gauge boson: $ X  \rightarrow (1) {\bar u}{\bar u}, (2) e^{-} d$.
Suppose that the branching ratio into the first channel with baryon number 
$B = - 2/3$ is $r$ and that of the second channel with baryon number 
$B = + 1/3$ is $1-r$. Suppose in addition that the branching ratio
for ${\bar X}$ into $({\bar 1}) uu$ with baryon number 
$B = + 2/3$ is ${\bar r}$ and
into $({\bar 2}) e^+ {\bar d}$ with baryon
number $B = - 1/3$ is $1- {\bar r}$.
Though the total decay rates of $X$ and ${\bar X}$ (normalized to unity)
are equal as required by CPT invariance, the differences in the individual
branching ratios signify a violation of C and CP conservation.

 Denote the parity (P)
of the states (1) and (2) by $\uparrow$ or 
$\downarrow$, then we have the following transformation properties:
\beq
\begin{array}{rccc}
{\rm Under~CPT:} \qquad & \Gamma(X \rightarrow 1 \uparrow) & = &
\Gamma(  {\bar 1} \downarrow \rightarrow {\bar X}) \\
{\rm Under~CP:} \qquad & \Gamma(X \rightarrow 1 \uparrow) & = &
\Gamma({\bar X} \rightarrow {\bar 1} \downarrow) \\
{\rm Under~C:} \qquad & \Gamma(X \rightarrow 1 \uparrow) & = &
\Gamma( {\bar X} \rightarrow {\bar 1} \uparrow ) .
\end{array}
\label{sym}
\eeq
We can now denote 
\begin{eqnarray}
r = \Gamma(X \rightarrow 1 \uparrow) + \Gamma(X \rightarrow 1 \downarrow) \\
{\bar r} = \Gamma( {\bar X} \rightarrow {\bar 1} \uparrow )
+ \Gamma({\bar X} \rightarrow {\bar 1} \downarrow) .
\end{eqnarray}
The total baryon number produced by an $X$, ${\bar X}$ decay
is then 
\begin{eqnarray}
\Delta B & = & -{2 \over 3} r + {1 \over 3} (1 - r) + {2 \over 3} {\bar r}
- {1 \over 3} (1 - {\bar r}) \nonumber \\
& = & {\bar r} - r = \Gamma( {\bar X} \rightarrow {\bar 1} \uparrow )
+ \Gamma({\bar X} \rightarrow {\bar 1} \downarrow) - 
\Gamma(X \rightarrow 1 \uparrow) - \Gamma(X \rightarrow 1 \downarrow) .
\end{eqnarray}
One sees clearly therefore, that from Eqs.~(\ref{sym}) if {\em either}
C {\em or} CP are good symmetries, $\Delta B = 0$.

In the out-of-equilibrium decay scenario~\cite{ww}, the total baryon asymmetry
produced is proportional to $\Delta B = ({\bar r} - r)$.
If decays occur out of equilibrium, then at the time of decay
$n_X \approx n_\gamma$ at $T < M_X$. We then have 
\beq
{n_B \over s} = {(\Delta B) n_X \over s} \sim 
{(\Delta B) n_X \over N(T) n_\gamma} \sim 10^{-2}(\Delta B) .
\label{nbmax}
\eeq
The schematic view presented above can be extended to a complete
calculation given a specific model~\cite{fot,kw}, see also
Ref.~\cite{kt} for reviews. 

\begin{figure}
\centering\includegraphics[width=.8\linewidth,angle=-0]{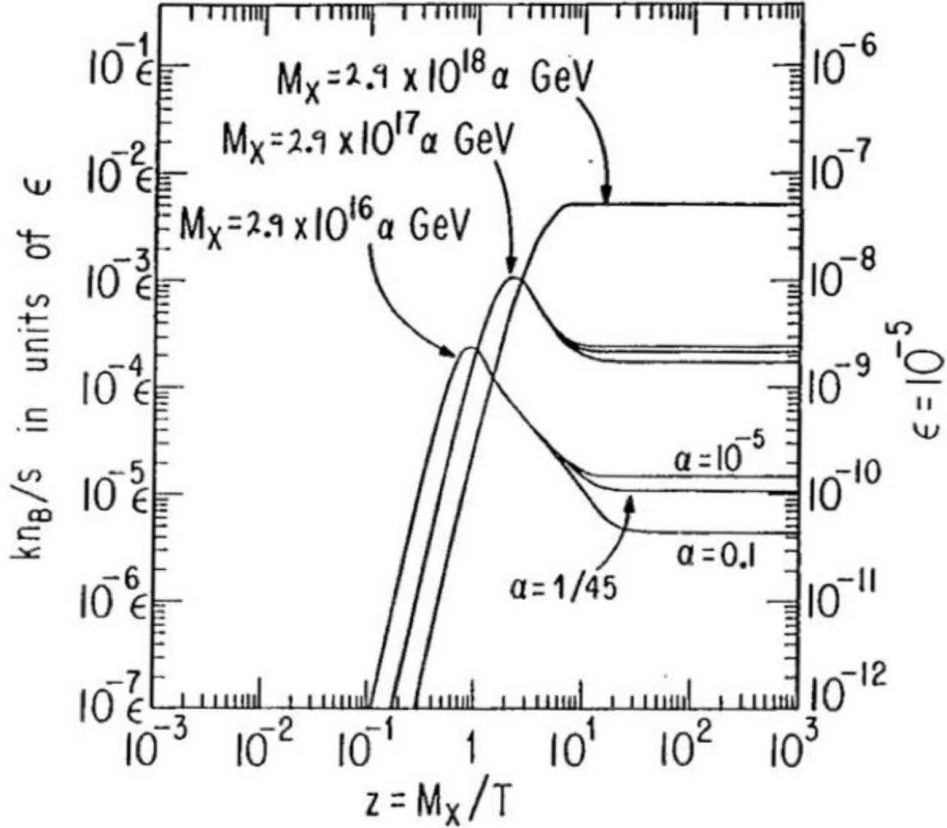}
\caption{The time evolution of the baryon asymmetry 
with $B = L = 0$ initially}
\label{bevol}
\end{figure}

The time evolution 
for the generation of a baryon asymmetry is shown in Fig.~\ref{bevol}.  As one
can see, for large values of $M_X$, i.e., values which satisfy the lower
limit given in Eq.~(\ref{mxmin}), the maximal value for the baryon asymmetry
$n_B/s \sim 10^{-2} \epsilon$ is achieved.  This confirms numerically
the original out-of-equilibrium decay scenario~\cite{ww}.  For smaller
values of $M_X$ an asymmetry is still produced which, 
however, is smaller due to  partial equilibrium
maintained by inverse decays.  The growth of the 
asymmetry 
as a function of time is now damped, and it reaches its final value when
inverse decays freeze out.
Finally, by studying different initial conditions, one can show that the result for the final baryon
asymmetry is in fact largely independent of the initial baryon asymmetry.

\begin{figure}
\centering\includegraphics[width=.8\linewidth]{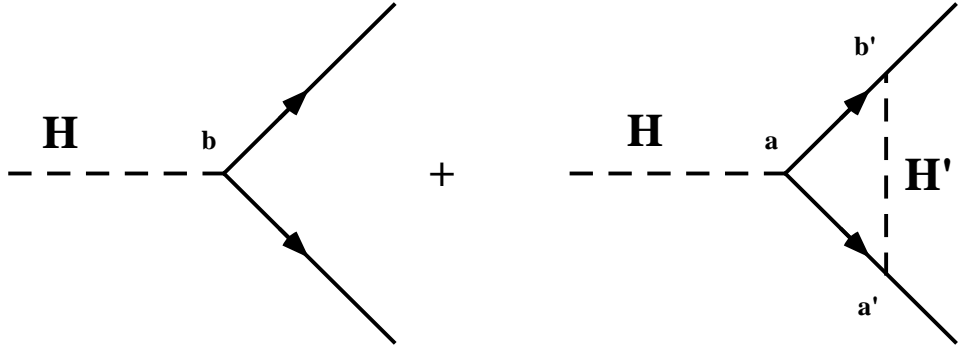}
\caption{One-loop contribution to the C and CP 
violation with two Higgs five-plets}
\label{hdecay}
\end{figure}

From Eq.~(\ref{nbmax}) it is clear that a complete calculation of
$n_B/s$ will require a calculation of the CP violation in the decays
(summed over parities) which we can parametrize by
\beq
\epsilon = {\bar r} - r = {\Gamma({\bar X} \rightarrow {\bar 1})
- \Gamma(X \rightarrow 1) \over \Gamma({\bar X} \rightarrow {\bar 1})
+ \Gamma(X \rightarrow 1)} \sim {{\rm Im} \Gamma \over {\rm Re} \Gamma}\ .
\eeq
At the tree level, as one can see, $\Gamma(X \rightarrow 1) \propto
g_5^\dagger g_5$ is real and there is no C or CP violation.
At the one loop level for gauge boson decay, 
there is also no net contribution to $\epsilon$, and we must turn to Higgs decay.
At least two Higgs five-plets are required to generate sufficient
C and CP violation.  With two five-plets,
$H$ and $H^\prime$, the interference of diagrams of the type in Fig.~\ref{hdecay}
will yield a non-vanishing $\epsilon$~\cite{yc},
\beq
\epsilon \propto {\rm Im} ({a^\prime}^\dagger a b^\prime b^\dagger) \ne 0\ ,
\eeq
if the couplings $a \ne a^\prime$ and $b \ne b^\prime$.

There are of course many alternative methods to generate
the baryon asymmetry, though each makes use of the same three ingredients.
For example, a supersymmetric mechanism proposed by 
Affleck and Dine~\cite{21} makes use of flat directions in the 
scalar potential.  There are many such flat directions, and
some of these yield an non-vanishing expectation value to 
GUT baryon number violating operators.  Supersymmetry breaking 
perturbs the flatness, leading to the cosmological evolution of the scalar
which oscillates about the global (charge and colour conserving) minimum of the potential.
Baryon number is stored in these oscillations and a net asymmetry is produced
as these fields decay.

Another mechanism to generate the baryon asymmetry employs
the heavy right-handed neutrinos used in the see-saw mechanism to 
generate neutrino masses~\cite{seesaw}.
The simplest of such mechanisms
is based on the decay of a right-handed neutrino-like state~\cite{20}. 
This mechanism
is certainly novel in that it does not require grand unification at all.
By simply adding to the Lagrangian a Dirac and Majorana mass term
 for a new right-handed neutrino state, 
\beq
{\cal L} \ni M\nu^c\nu^c + \lambda H L \nu^c ,
\eeq
the out-of-equilibrium decays $\nu^c \rightarrow L +  H^*$
 and  $\nu^c \rightarrow L^* + H$ will generate a non-zero 
lepton number $L \neq 0$. The out-out-equilibrium condition
for these decays translates to $10^{-3} \lambda^2 M_{\rm P} < M$
and $M$ could be as low as $O(10)$ TeV.
(Note that once again in order to 
have a non-vanishing contribution to the C and CP violation
in this process at 1-loop, at least two flavours of $\nu^c$ are required.
For the generation of masses of all three neutrino flavors,
three flavours of $\nu^c$ are required.)
 Electroweak sphaleron effects~\cite{6} can transfer this lepton asymmetry into a baryon
 asymmetry.
If sphalerons are in equilibrium, the baryon number can be expressed in terms
of $B-L$
\beq
B = {28 \over 79} \left( B - L \right) .
\label{2879}
\eeq
In the absence of a primordial $B-L$ asymmetry, 
the baryon number is erased by equilibrium processes.
Right-handed neutrinos produce a net lepton asymmetry and 
hence a net $B-L$ yielding the final baryon asymmetry given by Eq.~(\ref{2879}).

\section{Lecture 3: Big Bang Nucleosynthesis}

The standard model~\cite{wssok} of Big Bang Nucleosynthesis (BBN)
is based on the relatively simple idea of including an extended nuclear
network into a homogeneous and isotropic cosmology.  Apart from the
input nuclear cross sections, the theory contains only a single parameter,
namely the baryon-to-photon ratio,
$\eta$, and even that has been fixed by WMAP~\cite{wmap}. 
The theory then allows one to make
predictions (with well-defined uncertainties) of the abundances of the
light elements, D,
\he3, \he4, and \li7~\cite{cfo3}.

Conditions for the synthesis of the light elements were attained in the
early Universe at temperatures  $T \ga $ 1 MeV.  In the early Universe,
the energy density was dominated by radiation with
\begin{equation}
\rho = {\pi^2 \over 30} ( 2 + {7 \over 2} + {7 \over 4}N_\nu) T^4 ,
\label{rho}
\end{equation}
from the contributions of photons, electrons and positrons, and $N_\nu$
neutrino flavours (at higher temperatures, other particle degrees of
freedom should be included as well). At these temperatures, weak
interaction rates were in equilibrium. In particular, the processes
\begin{eqnarray}
n + e^+ & \leftrightarrow  & p + {\bar \nu_e} \nonumber \\
n + \nu_e & \leftrightarrow  & p + e^- \nonumber \\
n  & \leftrightarrow  & p + e^- + {\bar \nu_e} 
\label{beta}
\end{eqnarray}
fix the ratio of
number densities of neutrons to protons. At $T \gg 1$ MeV, $(n/p) \simeq
1$. 

As we have seen in the case of neutrino interactions, 
the weak interactions do not remain in equilibrium at lower temperatures.
Freeze-out occurs when the weak interaction rate $\Gamma_{wk} \sim G_F^2
T^5$ falls below the expansion rate which is given by the Hubble
parameter $H \sim T^2/M_{\rm P}$. The $\beta$-interactions in Eq.~(\ref{beta}) freeze out at about 0.8 MeV.
 As the temperature falls
and approaches the point where the weak interaction rates are no longer
fast enough to maintain equilibrium, the neutron-to-proton ratio is given
approximately by the Boltzmann factor,
$(n/p)
\simeq e^{-\Delta m/T} \sim 1/5$, where $\Delta m$ is the neutron--proton
mass difference. After freeze-out, free neutron decays drop the ratio
slightly to about 1/7 before nucleosynthesis begins. 
A useful semi-analytic description
of freeze-out has been given~\cite{bbf,muk}.

The nucleosynthesis chain begins with the formation of deuterium
by the process, $p+n \rightarrow$ D $+~\gamma$.
However, because of the large number of photons relative to nucleons,
$\eta^{-1} = n_\gamma/n_B \sim 10^{10}$, deuterium production is delayed
past the point where the temperature has fallen below the deuterium
binding energy, $E_B = 2.2$ MeV (the average photon energy in a blackbody
is ${\bar E}_\gamma \simeq 2.7\,T$).  This is because there are many
photons in the exponential tail of the photon energy distribution with
energies $E > E_B$ despite the fact that the temperature or ${\bar
E}_\gamma$ is less than $E_B$.  The degree to which deuterium production
is delayed can be found by comparing the qualitative expressions for the
deuterium production and destruction rates,
\begin{eqnarray}
\Gamma_p & \approx & n_B \sigma v \\ \nonumber
\Gamma_d & \approx & n_\gamma \sigma v e^{-E_B/T} .
\end{eqnarray}
When the quantity $\eta^{-1}
{\rm exp}(-E_B/T)
\sim 1$, the rate for  deuterium destruction (D $+ ~\gamma
\rightarrow p + n$) finally falls below the deuterium
production rate and the nuclear chain begins at a temperature
$T \sim 0.1 {\rm ~MeV}$.

\begin{figure}
\centering\includegraphics[width=.8\linewidth,angle=270]{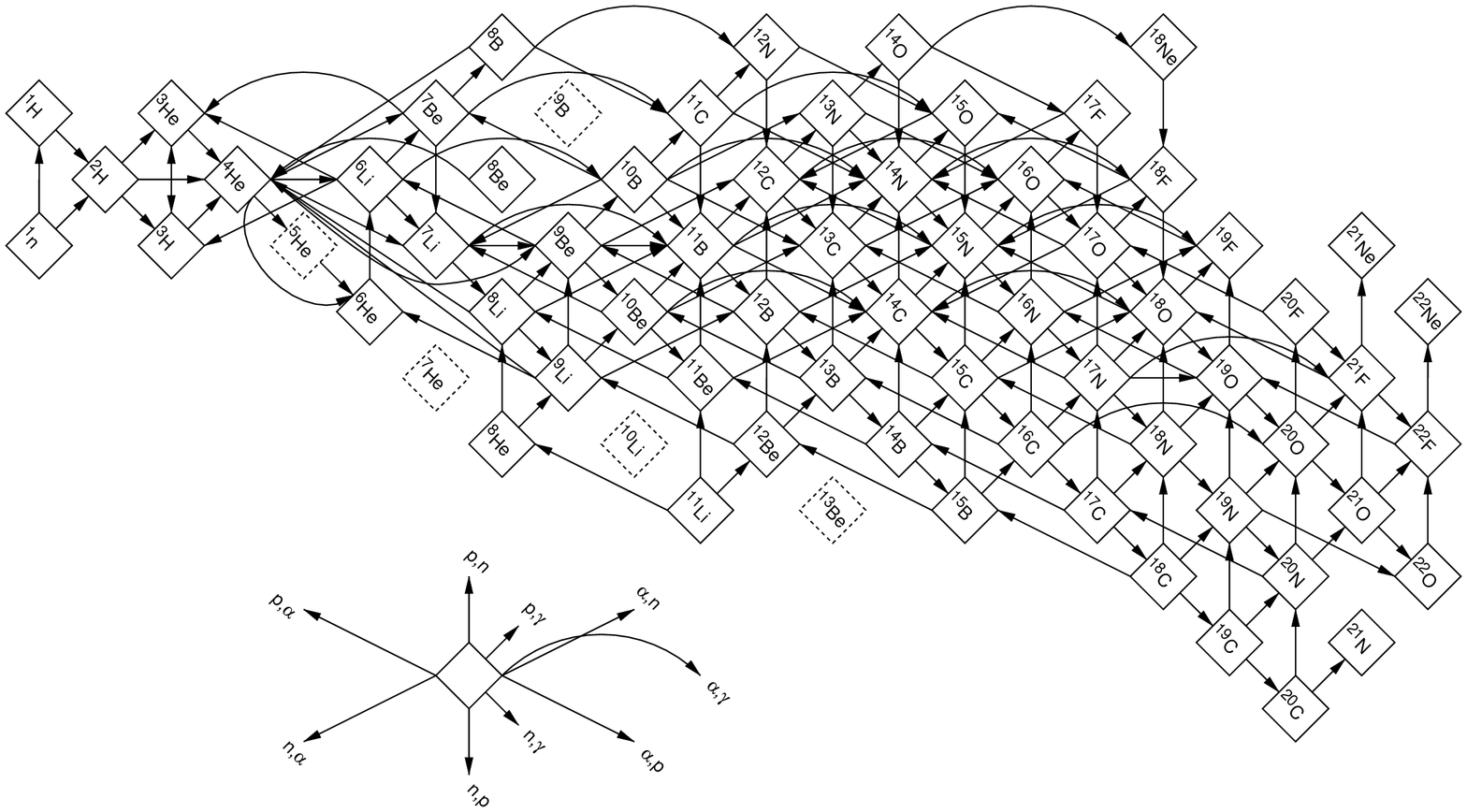}
\caption{The nuclear network used in BBN calculations}
\label{net}
\end{figure}

The dominant product of Big Bang nucleosynthesis is \he4 and its
abundance is very sensitive to the
$(n/p)$ ratio
\begin{equation}
Y_p = {2(n/p) \over \left[ 1 + (n/p) \right]} \approx 0.25 ,
\label{ynp}
\end{equation}
i.e., an
abundance of close to 25\% by mass. Lesser amounts of the other light
elements are produced: D and \he3 at the level of about $10^{-5}$ by
number,  and \li7 at the level of $10^{-10}$ by number. 
The gap at $A=8$ prevents the production of other isotopes in any
significant quantity. The nuclear chain is shown in Fig.~\ref{net}.

\begin{figure}[htbp]
\begin{center}
\includegraphics[width=20pc]{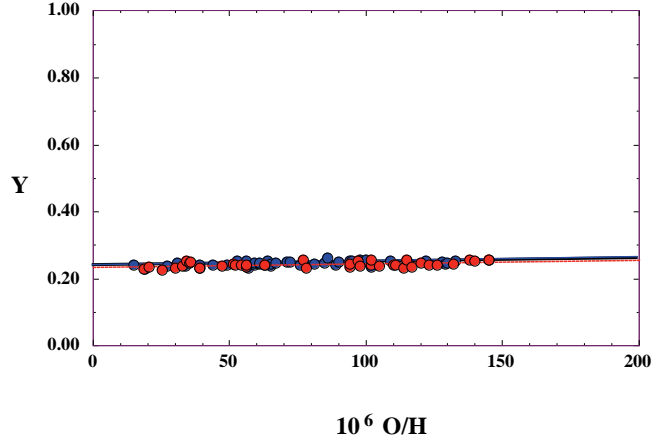}
\end{center}
\caption{{The \he4 mass fraction as determined in extragalactic H~II
regions as a function of O/H}}
\label{primhe4}
\end{figure}

Historically, BBN as a theory explaining the observed element abundances
was nearly abandoned due its inability to explain {\em all} element
abundances. Subsequently, stellar nucleosynthesis became the leading
theory for element production~\cite{bbfh}.  However, two key questions
persisted. 1) The abundance of \he4 as a function of metallicity is
nearly flat and no abundances are observed to be below about 23\% as
exaggerated in Fig.~\ref{primhe4}. In particular, even
in systems in which an element such as oxygen which traces stellar
activity is observed at extremely low values (compared with the solar
value of O/H $\approx 4.9 \times10^{-4}$), the \he4 abundance is nearly
constant.  This is very different from all other element abundances (with
the exception of \li7 as we shall see below).  For example, in Fig.~\ref{no}, the N/H vs. O/H correlation is shown~\cite{fdo}.  As one can clearly see,
the abundance of N/H goes to 0, as O/H goes to 0, indicating a stellar
source for nitrogen. 
2) Stellar sources cannot produce the observed abundance of D/H.
Indeed, stars destroy deuterium and no astrophysical site is known
for the production of significant amounts of deuterium~\cite{rafs}.
Thus we are led back to BBN for the origins of D, \he3, \he4, and \li7.

\begin{figure}[htbp]
\begin{center}
\includegraphics[width=20pc]{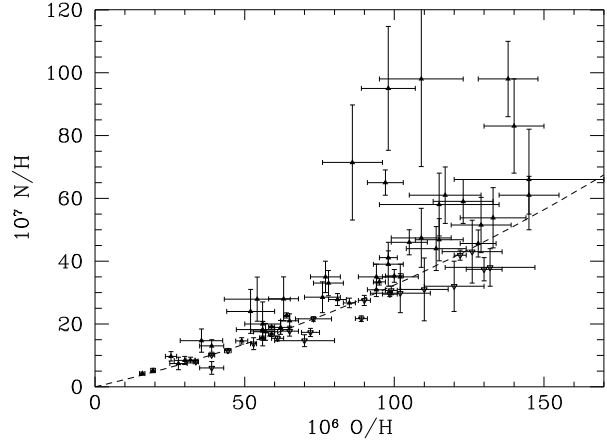}
\end{center}
\caption{{The nitrogen and oxygen abundances in the same extragalactic
H~II regions with observed \he4 shown in Fig.~\protect\ref{primhe4}}}
\label{no}
\end{figure}

\subsection{Abundance predictions}

Because standard BBN theory rests upon the 
Standard Model of particle physics, 
the electroweak aspects of the calculation
are very well-determined
and do not introduce an appreciable uncertainty.
Instead, the major uncertainties come from the
thermonuclear reaction rates.
There are 11 key strong
rates (as well as the neutron lifetime) which
dominate the uncertainty budget~\cite{cfo1,kr,nb,cvcr01}.
In contrast to the situation for much of
stellar nucleosynthesis, BBN occurs at high
enough temperatures that laboratory data
exist at and even below the relevant energies, so
that no extrapolation is needed.
 Monte Carlo techniques~\cite{cfo1,kr} are used to determine the best-fit abundances,
and their uncertainties, at each $\eta$.

Recently the
input nuclear data have been carefully reassessed~\cite{nb,cfo1,cvcr01,nacre,coc},
leading to improved precision in the abundance predictions. 
In addition, polynomial
fits to the predicted abundances and the error correlation matrix have
been given~\cite{flsv,cyburt}. 
The NACRE Collaboration presented an updated nuclear
compilation~\cite{nacre}.  
For example, notable improvements include a reduction in the uncertainty
in the rate for \he3$(n,p)$T from 10\%~\cite{skm} to 3.5\% and
for T$(\alpha, \gamma)$\li7 from $\sim 23$--$30\%$~\cite{skm} to $\sim 4\%$.
Since then, new data and techniques have become available, motivating
new compilations.  Within the last year, several new BBN compilations
have been presented~\cite{cyburt,coc2,cuoco,cfo5}.

\begin{figure}
\centering\includegraphics[width=.9\linewidth]{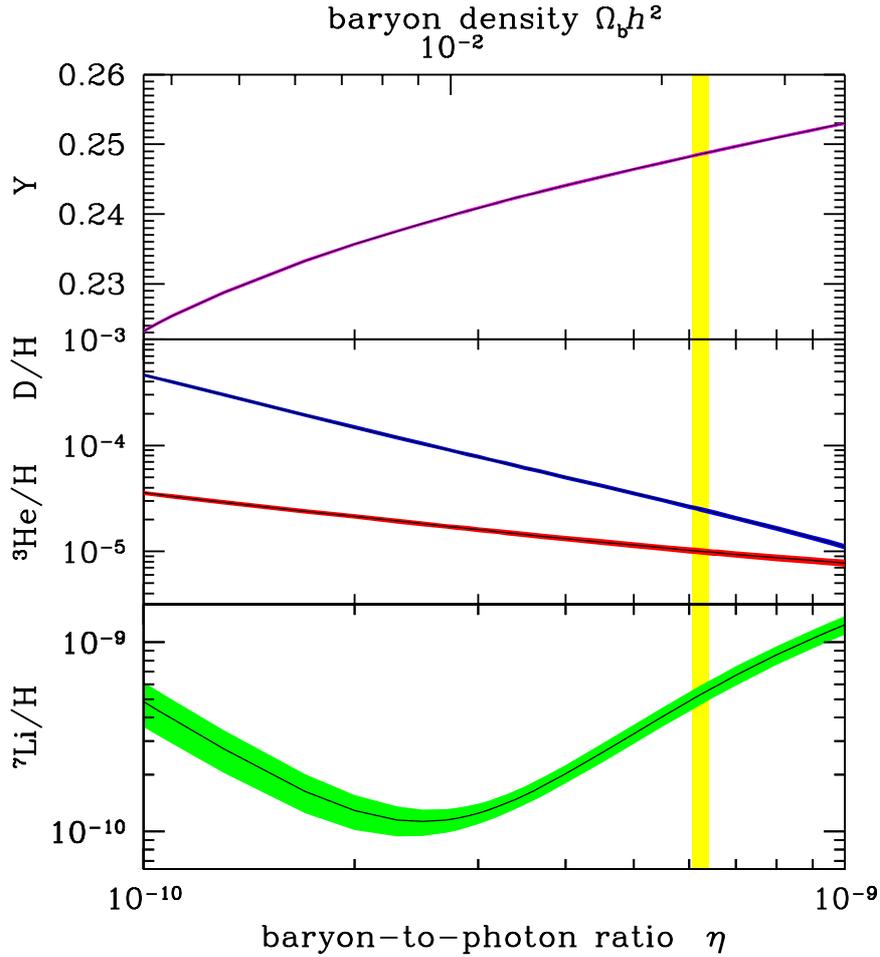}
\caption{{The predictions of standard BBN~\cite{cfo5} showing the
primordial abundances as a function of the
baryon-to-photon ratio $\eta$.  Abundances are quantified as
ratios to hydrogen, except for \he4 which is given in
baryonic mass fraction $Y_p = \rho_{\rm He}/\rho_B$. 
The lines give the mean values, and the surrounding
bands give the $1\sigma$ uncertainties.
}}
\label{cfo5}
\end{figure}

The light element abundances are shown in Fig.~\ref{cfo5}
 as a function of $\eta$~\cite{cfo5}. 
 The plot shows the
abundance of \he4 by mass $Y$ and the abundances of the other three
isotopes by number.  The curves indicate the central predictions from
BBN, while the bands correspond to the uncertainty in the predicted
abundances.
The uncertainty range in \he4 reflects primarily the 1$\sigma$ uncertainty in
the neutron lifetime.

In standard BBN with $N_\nu = 3$, the only free parameter is the
density of baryons (strictly speaking, nucleons),
which sets the rates of the strong reactions.
Because standard BBN is a one-parameter theory, 
any abundance measurement determines $\eta$, while additional measurements
overconstrain the theory and thereby provide a consistency check.
BBN has thus historically been the premier means of determining
the cosmic baryon density.

The release of the first-year WMAP results 
on the anisotropy spectrum of the CMB were a landmark event
for all of cosmology, but particularly for BBN. 
As discussed above, the value of $\eta$ has been
fixed by CMB measurements as given by Eq.~(\ref{etas}). 
Thus, within the context
of the Standard Model, 
BBN becomes a zero-parameter theory, and the light element
predictions are completely determined to within the uncertainties in
$\eta$ and the BBN theoretical errors. Comparison with light
element observations then can be used to restate the test of BBN--CMB
consistency, or to turn the problem around and test the astrophysics
of post-BBN light element evolution~\cite{cfo2}. 

\subsection{Light element observations and comparison with theory} 

BBN theory predicts the abundances of D, \he3, \he4, and
\li7, which are essentially determined at $t\sim180$~s. Abundances are,
however, observed at much later epochs, after stellar nucleosynthesis
has commenced. The ejecta from stellar processing
can alter the light element abundances from their primordial values,
but also produce heavy elements such as C, N, O, and Fe
(`metals'). Thus one seeks astrophysical sites with low metal
abundances, in order to measure light element abundances which are
closer to primordial.
For all of the light elements, 
systematic errors are an important and often dominant
limitation to the precision of the primordial abundances.

\subsubsection{{\rm  \bf D/H}}

In recent years, high-resolution spectra have revealed the presence of
D in high-redshift, low-metallicity quasar absorption systems (QAS),
via its isotope-shifted Lyman-$\alpha$ absorption. These are the first measurements of light
element abundances at cosmological distances. It is believed that
there are no astrophysical sources of deuterium~\cite{rafs}, so any
measurement of D/H provides a lower limit to primordial D/H and thus
an upper limit on $\eta$.   
Recent observations by FUSE show a wide dispersion in the
deuterium abundance in local gas seen via its absorption,
$({\rm D/H})_{\rm local~gas} = (0.5-2.2)\times10^{-5}$~\cite{fuse}.
This surprisingly large spread, taken together with the positive correlation
of D/H with temperature and metallicity along various sightlines,
led~\cite{fuse} to suggestions that deuterium may suffer significant
and preferential depletion onto dust grains.  In this case
the true local interstellar D/H value would lie at the upper limit
of the observed values, giving
${(\rm D/H)}_{\rm ISM} \ga (2.31 \pm 0.24) \times 10^{-5}$.
However, extracting a primordial deuterium value requires a Galactic
chemical evolution model (e.g., Ref.~\cite{galdeut}), whose model dependences yield
uncertainties in the determination of the primordial deuterium abundance.
Many of these models do not predict significant D/H
depletion~\cite{fie} at high redshift, and in this case the high-redshift 
measurements are expected to recover the primordial deuterium abundance.

\begin{figure}
\centering\includegraphics[width=.6\linewidth]{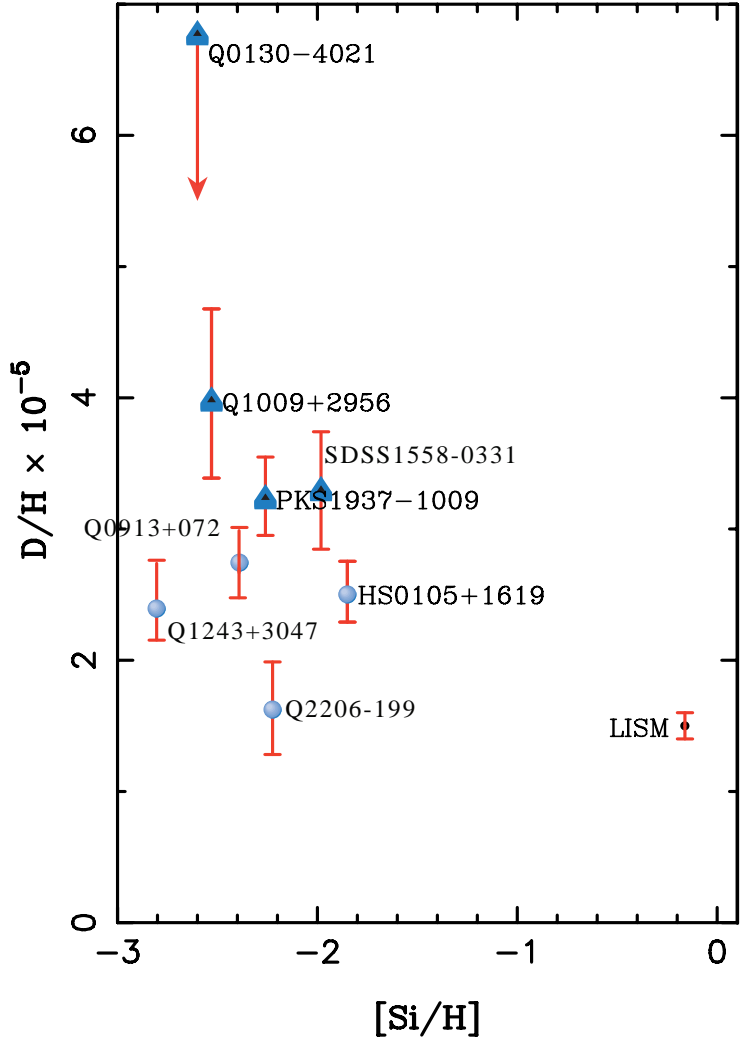}
\caption{{D/H abundances shown as a function of [Si/H].
Labels denote the background quasi-stellar objects~(QSO), except for
the local interstellar value (LISM;~\cite{fuse}).}}
\label{D}
\end{figure}

The deuterium abundance at low metallicity has been measured in
several quasar absorption systems~\cite{pettini}.  The weighted mean value of
the seven systems with reliable abundance determinations is $\log$ D/H
$= -4.55 \pm 0.03$ where the error includes a scale factor of 1.72
and corresponds to D/H = $(2.82 \pm 0.21) \times
10^{-5}$. These are shown in Fig.~\ref{D}. 
Since the D/H shows considerable scatter
it is likely that systematic errors dominate the uncertainties.
In this case it may be more
appropriate to derive the uncertainty using sample variance (see, for example, Ref.~\cite{cfo1}) which gives 
a more conservative range $\log$ D/H $= -4.55 \pm 0.08$ or D/H =
$(2.82 \pm 0.53) \times 10^{-5}$.

Using the WMAP value for the baryon density~(\ref{etas}),
the primordial D/H abundance is predicted to be~\cite{cfo5}
\beq
({\rm D/H})_p = (2.49 \pm 0.17) \times 10^{-5} .
\label{dpred}
\eeq
As one can see from Fig.~\ref{cfo5}, this is in good
agreement with the average of the seven best determined quasar absorption system 
abundances noted above, particularly when systematic uncertainties are taken into account.

\subsubsection{\he4}

We observe \he4 in clouds of ionized hydrogen (H~II regions), the most
metal-poor of which are in dwarf galaxies. There is
now a large body of data on \he4 and CNO in these systems~\cite{iz}. These data confirm that the small stellar
contribution to helium is positively correlated with metal production.
Recently a careful study of the systematic uncertainties in \he4,
particularly the role of underlying absorption, has led to a higher value 
for the primordial abundance of \he4~\cite{os2}.
Using a subset of the highest quality from the data of Izotov and Thuan~\cite{iz},
all of the physical parameters listed above including the \he4 abundance were
determined self-consistently with Monte Carlo methods~\cite{os}. 
The extrapolated \he4 abundance was determined to be 
$Y_p = 0.249 \pm 0.009$.  
Conservatively, it would be difficult at this time to exclude any value of
$Y_p$ inside the range 0.232--0.258.

At the WMAP value for $\eta$, the \he4 abundance is predicted to be~\cite{cfo5}
\beq
\label{eq:Yp}
Y_p = 0.2486 \pm 0.0002 .
\eeq
This is in excellent agreement with the most recent analysis of the \he4 abundance~\cite{os2}. Note also that the large uncertainty ascribed to this value
indicates that while \he4 is certainly consistent with the WMAP 
determination of the baryon density, it does not provide for
a highly discriminatory test of the theory at this time.

\subsubsection{\li7}

The systems best suited for Li observations are metal-poor stars
in our Galaxy. Observations
have long shown~\cite{sp} that Li does not vary
significantly in Pop II stars with metallicities $\la1/30$ of solar
--- the `Spite plateau'. Precision data
suggest a small but significant correlation between Li and Fe~\cite{rnb} which can be understood as the result of Li production
from Galactic cosmic rays~\cite{van}. Extrapolating to
zero metallicity one arrives at a primordial value~\cite{rbofn}
${\rm Li/H}|_p =
 (1.23^{+0.34}_{-0.16}) \times 10^{-10}$.
 
 \begin{figure}[h] 
\begin{center}
\includegraphics[width=25pc]{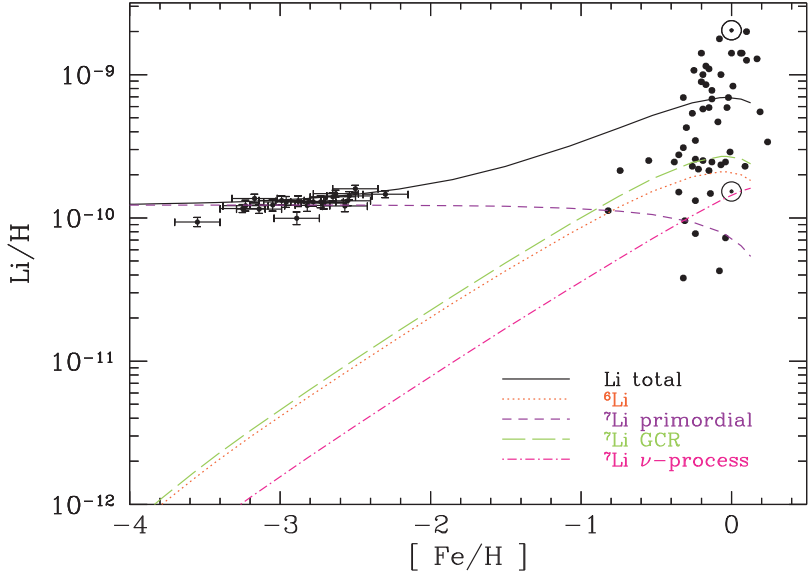}
\end{center}
\caption{Contributions to the total predicted lithium abundance from
the adopted Galactic chemical evolution model of Ref.~\protect\cite{fo99}, compared with low
metallicity  stars  and a sample of high
metallicity stars. The solid curve is the sum of all  components.}
\label{li2fig}
\end{figure}

Figure~\ref{li2fig} shows the different Li
components for a model with (\li7/H)$_p = 1.23 \times 10^{-10}$
as a function of the iron abundance expressed as the log of Fe/H relative to the solar value. 
The linear slope produced by the model is independent of the input
primordial value. The model~\cite{fo99} includes, in addition to primordial \li7, lithium produced in
Galactic cosmic-ray nucleosynthesis (primarily $\alpha + \alpha$ fusion),
and \li7 produced by the $\nu$-process during type II supernovae. As one
can see, these processes are not sufficient to  reproduce the population
I abundance of \li7 (at near solar [Fe/H] $\la 0$), and additional production sources are needed.

A recent reanalysis of the
$\he3(\alpha,\gamma)\be7$
reaction, which is the most important \li7 production process in BBN,
was considered in detail in Ref.~\cite{cd}. When the new rate is used a
high \li7 abundance is found~\cite{cfo5} at the WMAP value of $\eta$
\beq
 \li7/{\rm H} = (5.24^{+0.71}_{-0.62}) \times 10^{-10} .
 \label{bbn7}
 \eeq
 This represents a 23\% increase in \li7 over previous 
calculations~\cite{cyburt}.  The increase is primarily due to an increase in the
\he3($\alpha,\gamma$)\be7 cross section. Newer data~\cite{cd} implies 
17\% increase in this reaction leading to a 16\% increase in \li7.
In addition, the 1.5\% increase in $\eta$ from the 3-year to 5-year
WMAP data~\cite{wmap} leads to a 3\% increase in \li7 and finally
another 1\% increase is due to updated $pn$ rates. 
In addition, the uncertainty in the BBN \li7 abundance is roughly a factor of 2 times
smaller than previous determinations.
 This value for primordial \li7 is in clear contradiction with most estimates of the 
primordial Li abundance. 
 Several attempts at explaining this
discrepancy by adjusting some of the key nuclear rates have proved
unsuccessful~\cite{coc,coc3,cfo4}.
 
 An important source for systematic error lies in the derived effective
temperature of the star.  [Li] $ = \log (\li7/{\rm H}) + 12$ is very
sensitive to the temperature, with $\partial {\rm [Li]}/\partial
T_{\rm eff} \simeq$ 0.065 -- 0.08.  Unfortunately there is no 
standard for determining effective temperatures, and, for a given star,
there is considerable range depending on the method used.  This spread
in temperatures was made manifest in the recent work of Melendez and
Ramirez~\cite{mr} using the infra-red flux method (IRFM) which showed
differences for very low metallicities ([Fe/H] $<$ -3) by as much as
500 K, with typical differences of $\sim 200$ K with respect to that
of Ref.~\cite{rnb}.  As a consequence the derived \li7 abundance was
significantly higher with ${\rm Li/H}|_p = (2.34 \pm 0.32) \times
10^{-10}$~\cite{mr,fov}.

Recently a dedicated set of observations were performed with the specific goal
of determining the effective temperature in metal-poor stars~\cite{hos}.  Using a large set of Fe I excitation lines ($\sim 100$
lines per star), the Boltzmann equation was used with the excitation
energies, $\chi_i$ to determine the temperature through the
distribution of excited levels.  Again, there was no evidence for the
high temperatures reported in Ref.~\cite{mr}, rather, temperatures were
found to be consistent with previous determinations.  The mean \li7
abundance found in Ref.~\cite{hos} was ${\rm Li/H} = (1.3 - 1.4 \pm 0.2)
\times 10^{-10}$, consistent with the bulk of prior abundance
determinations.

There are of course other possible sources of systematic uncertainty
in the \li7 abundance.  It is possible that some of the surface \li7
has been depleted if the outer layers of the stars have been
transported deep enough into the interior, and/or mixed with material
from the hot interior; this may occur due to convection, rotational
mixing, or diffusion.  Estimates for possible depletion factors are in
the range $\sim$~0.2--0.4~dex, i.e., factors of 1.6--2.5~\cite{dep}. Recent attempts to deplete
the \li7 abundance through diffusion introduce a source of
turbulence tuned to fit the abundances of heavy elements in NGC6397~\cite{diff}.  Once parameters are set, the degree of lithium depletion
becomes a prediction of the model. For this cluster, a depletion factor of 
0.25 dex is found, i.e., a factor of 1.8. 
Note that while this depletion factor would bring the previous BBN
result of \li7/H = $4.26 \times 10^{-10}$~\cite{cyburt} to a value close the
value observed in that cluster~\cite{bon1} ($2.2 \times 10^{-10}$), a larger depletion factor is needed
with the new BBN value for \li7 given above.
It is also not clear whether this mechanism will work for the
wide range of stellar parameters seen in the field. As noted above,
the Li data show a negligible intrinsic spread in Li.  Any mechanism
which reduces significantly the abundance of \li7 must do so uniformly
over a wide range of stellar parameters (temperature, surface gravity,
metallicity, rotational velocity, etc.). 

It is also possible that the lithium discrepancy is a sign
of new physics beyond the Standard Model.  One possibility is the
cosmological variation of the fine structure constant. Varying
$\alpha$ would induce a variation in the deuterium binding energy and
could yield a decrease in the predicted abundance of \li7~\cite{vary}.
A potential solution to both lithium problems is particle decay after
BBN which could lower the \li7 abundance (and produce some \li6 as well)~\cite{jed04}. This has been investigated in the framework of the
constrained minimal supersymmetric Standard Model if the lightest
supersymmetric particle is assumed to be the gravitino~\cite{susy} and
indeed, some models have been found which accomplish these goals~\cite{susy2}.

 \subsubsection{\he3}
 
Since \he3 is also quite sensitive to the baryon density, one might hope
that it too could be used as a baryometer.  Observations of H~II
regions in our own Galaxy yield values of the \he3/H ratio that are compatible with
calculations of the primordial value~\cite{bania,vofc}.
However, the extrapolation from the
observations to a primordial abundance is complicated by the unknown chemical
evolution of \he3.  Indeed, one does not even know whether \he3/H is
increasing or decreasing with cosmic time.  Thus a primordial
extrapolation yields only an order-of-magnitude range of allowable
values of \he3/H~\cite{foscv}.

\subsection{Beyond the Standard Model}

Given the simple physics underlying BBN, it is remarkable that it
still provides one of the most effective tests for physics beyond the Standard Model. 
Limits on particle physics beyond the Standard Model come mainly from
the observational bounds on the \he4 abundance.
As discussed earlier, the neutron-to-proton
ratio is fixed by its equilibrium value at the freeze-out of 
the weak interaction rates at a temperature $T_f \sim 1$ MeV modulo the
occasional free neutron decay.  Furthermore, freeze-out is determined by
the competition between the weak interaction rates and the expansion rate
of the Universe
\begin{equation}
{G_F}^2 {T_f}^5 \sim \Gamma_{\rm weak}(T_f) = H(T_f) \sim \sqrt{G_N N} {T_f}^2 .
 \label{comp}
\end{equation}
In the
Standard Model, the number of relativistic particle species at 1~MeV
is $N = \frac{11}{2} + {7\over4}N_\nu$. The presence
of additional neutrino flavours (or any other relativistic species) at 
the time of nucleosynthesis increases the overall energy density
of the Universe and hence the expansion rate leading to a larger 
value of $T_f$, $(n/p)$, and ultimately $Y_p$.  Because of the
form of Eq.~(\ref{comp}) it is clear that just as one can place
limits~\cite{ssg} on $N_\nu$, any changes in the weak or gravitational
coupling constants can be similarly constrained (for a discussion see Ref.~\cite{co}).

\begin{figure}[t] 
\begin{center}
\includegraphics[width=20pc]{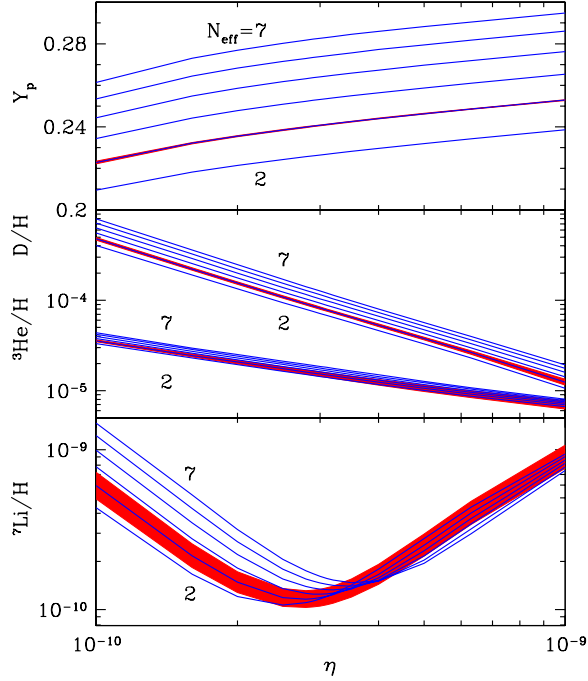}
\end{center}
 \vskip -.5in
\caption{BBN abundance predictions~\protect\cite{cfo2} as a function of
the baryon-to-photon ratio $\eta$, for $N_\nu =$~2--7.  The bands show
the $1\sigma$ error bars.  Note that for the isotopes other than Li, the
error bands are comparable in width to the thickness of the abundance
curve shown. All bands are centred on $N_\nu = 3$.}
\label{nnu1}
\end{figure}

\begin{figure}[t] 
\begin{center}
\includegraphics[width=0.6\linewidth,angle=270]{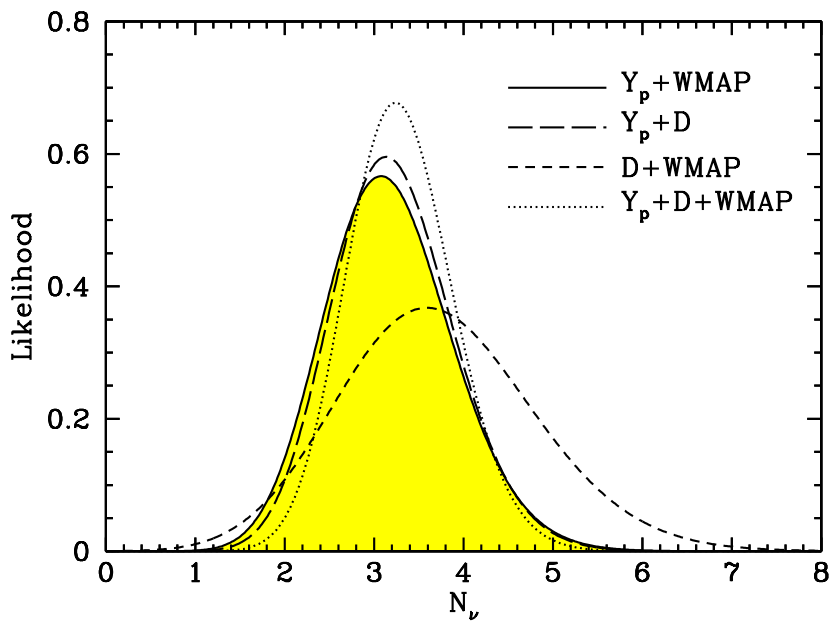}
\end{center}
\vskip -.5in
\caption{The likelihood distribution for $N_\nu$ based on the WMAP value of $\eta$ and 
$Y_p$ (shaded), WMAP and D/H (dashed), WMAP and both
$Y_p$ and D/H$_A$ (dotted).  Also shown is the result without imposing the WMAP
value for $\eta$ (long dashed).}
\label{fig:LN}
\end{figure}

 The helium curve in Fig.~\ref{cfo5} was computed taking
$N_\nu = 3$; the computed abundance scales as $\Delta Y_{BBN} \simeq
0.013 \Delta N_\nu$~\cite{bbf}. 
The dependence of the light element abundances on $N_\nu$
is shown in Fig.~\ref{nnu1}~\cite{cfo2}.
For a fixed value of $\eta = (6.14 \pm 0.25) \times 10^{-10}$ (slightly below the current 
WMAP value) and $Y_p = 0.249 \pm 0.009$, 
the likelihood distribution for $N_\nu$ is shown by the shaded region
in Fig.~\ref{fig:LN}~\cite{cfos}.  Also shown for comparison are the likelihood distribution
based the WMAP value of $\eta$ using D/H alone, $Y_p$ and D/H, and the
result based on BBN alone. Despite the increased uncertainty in the He abundance,
it still provides the strongest constraint on $N_\nu$.  D/H is nonetheless becoming
competitive in its ability to set limits on $N_\nu$. 

The 95 \% CL upper limits to $N_\nu$ are given in Table~\ref{tab:nu}.
In all cases the preferred values
for $N_\nu$ are consistent with $N_\nu$,
and in many cases are much closer to $N_\nu$ than
$1\sigma$.  This restates the overall consistency
among standard BBN theory, D and \he4 observations,
and CMB anisotropies. It also constrains departures
from this scenario.
The combined limit using BBN + light elements + CMB limit is~\cite{cfos}
\beq
\label{eq:Nnu}
2.67 \le N_\nu \le 3.85 
\eeq
at 68\% CL.

\begin{table}[ht]
\begin{center}\caption{This table shows constraints placed on $N_{\nu}$ and 
$\eta$ by various combinations of observations.  Shown are the 68\%
confidence limits determined by marginalizing the 2-D likelihood
distribution. Also shown are the 95\%
upper limits on $\delta N_{\nu}=N_{\nu}-3$, given that $\delta N_{\nu} > 0$.}
\label{tab:nu}
\medskip
\begin{tabular}{lc|c|c}
{\bf Observations} & \ \ \boldmath$\eta_{10}\equiv 10^{10}\eta$ \ \ & \boldmath$N_{\nu}$ &
\boldmath$\delta N_{\nu,max}$ \\[.1cm]
\hline
{\vrule height 15pt depth 0pt width 0pt}
$Y_p$ + D/H$_A$ & $5.94^{+0.56}_{-0.50}$ & $3.14^{+0.70}_{-0.65}$ & 1.59 \\[.1cm]
\hline
{\vrule height 15pt depth 0pt width 0pt}
$Y_p$ + $\eta_{CMB}$ & $6.14\pm0.25$ & $3.08^{+0.74}_{-0.68}$ & 1.63 \\[.1cm]
\hline
{\vrule height 15pt depth 0pt width 0pt}
D/H$_A$ + $\eta_{CMB}$ & $6.16\pm0.25$ & $3.59^{+1.14}_{-1.04}$ & 2.78 \\[.1cm]
\hline
{\vrule height 15pt depth 0pt width 0pt}
$Y_p$ + D/H$_A$ + $\eta_{CMB}$ & $6.10^{+0.24}_{-0.22}$ & $3.24^{+0.61}_{-0.57}$ & 1.44 \\[.1cm]
\hline
\end{tabular}
\end{center}
\end{table}

\section{Lecture 4: Dark Matter}

Evidence for dark matter in the Universe is available from a wide range of
observational data.  As discussed many times above, 
the analysis of the cosmic microwave background anisotropies leads to the conclusion
that the curvature of the Universe is close to zero indicating that the sum of the
fractions of critical density, $\Omega$,  in matter and a cosmological constant (or dark energy)
is very close to one~\cite{wmap}.  When combined with a variety of data including results from
the analysis of type Ia supernovae observations~\cite{sn1,sn2} and baryon acoustic oscillations~\cite{bao} one is led to the concordance model where $\Omega_m \sim 0.23$ and $\Omega_\Lambda
\sim 0.73$ with the remainder (leading to $\Omega_{tot} = 1$) in baryonic matter. 
This is in addition to the classic evidence from galactic rotation curves~\cite{rot}, which indicate
that nearly all spiral galaxies are embedded in a large galactic halo of dark matter
leading to rather constant rotational velocities at large distances from the centre of the galaxy
(in contrast to the expected $v^2 \sim 1/r$ behaviour in the absence of dark matter).
Other dramatic pieces of evidence can be found in combinations of X-ray observations
and weak lensing showing the superposition of dark matter (from lensing) and ordinary matter
from X-ray gas~\cite{witt} and from the separation of baryonic and dark matter
after the collision of two galaxies as seen in the Bullet cluster~\cite{clowe}.
For a more complete discussion see Ref.~\cite{otasi3}.

From the first column of Table~\ref{tab:wmap}, we can obtain the density of cold dark matter from
the difference between the total matter density and the baryon density~\cite{wmap}
\beq
\Omega_{CDM} h^2 = 0.1099 \pm 0.0062
\label{wmap}
\eeq
or a 2$\sigma$ range of 0.0975--0.1223 for $\Omega_{CDM} h^2$.

\subsection{Neutrinos}

\begin{figure}[t]
\centerline{
\includegraphics[width=0.8\textwidth]{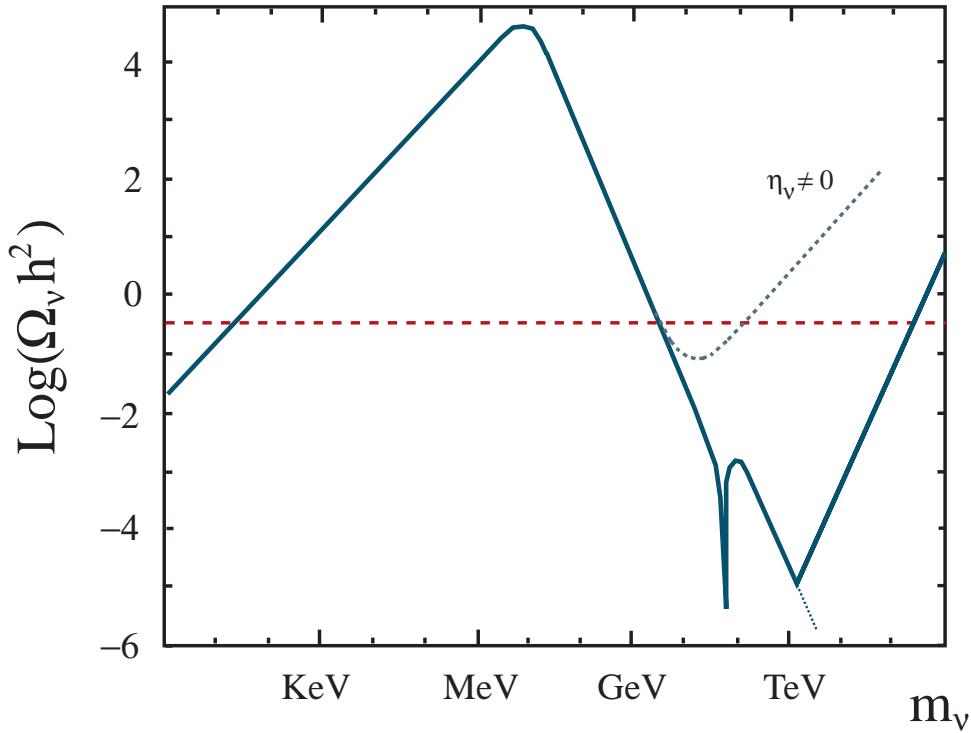}
}
\caption{{Summary plot~\protect\cite{kko} of the relic density of Dirac
neutrinos  (solid) including a possible neutrino asymmetry of $\eta_\nu =
5\times 10^{-11}$ (dotted)}}
\label{nu}
\end{figure}

Dark matter must be both long-lived or stable and electrically and colour neutral.
As a result, once baryons and neutrinos are eliminated as candidates,
one must look beyond the Standard Model. From WMAP, we already know that the
baryon density is far below the requisite amount in cold dark matter.
	Light neutrinos ($m \le 30 {\rm ~eV}$) are 
a long-time standard when it comes to
 non-baryonic dark matter~\cite{ss}.    Light neutrinos
 are, however, ruled out as a dominant form of dark matter because they 
 produce too much large scale structure~\cite{nu3}.
The energy of density of light neutrinos with $m_\nu \la 1$ MeV can be 
expressed at late times as
$  \rho_\nu  = \frac{3}{11} m_\nu  n_\gamma$.
Imposing the constraint $\Omega_\nu h^2 \la 0.12$, translates into a 
 strong constraint (upper bound) on Majorana
neutrino masses~\cite{cows}:
\begin{equation}
m_{\rm tot} =   \sum_\nu  m_\nu   \la 11 {\rm ~eV} ,
\label{ml1}
\end{equation}
where the sum runs over neutrino mass eigenstates. The limit for Dirac
neutrinos depends on the interactions of the right-handed states.  
The limit~(\ref{ml1}) and the
corresponding  initial rise in
$\Omega_\nu h^2$ as a function of $m_\nu$ is  displayed in
Fig.~\ref{nu}.
Much stronger limits on the sum of neutrino masses are
possible when combining the WMAP data with large scale structure surveys.
A typical limit is $m_{\rm tot} < 0.7$ eV or $\Omega_\nu h^2 < 0.0076$~\cite{wmap}.

The calculation of the relic density for neutrinos more massive than
$\sim 1$ MeV, is substantially more involved. The relic density is now
determined by the freeze-out of neutrino annihilations which occur at
$T \la m_\nu$, after annihilations have begun to seriously reduce their
number density~\cite{lw}.  For particles
which annihilate through approximate weak scale interactions, annihilations freeze out
when $T \sim m_\chi /20$.  

Based on the leptonic and invisible width of the $Z$ boson, 
experiments at LEP have determined that the number of neutrinos is 
$N_\nu = 2.994 \pm 0.012$~\cite{rpp}. Thus, LEP excludes additional 
neutrinos (with standard weak interactions) with masses $m_\nu 
\la 45$ GeV.  The mass density of ordinary heavy
neutrinos is found to be very small, $\Omega_\nu {h}^2 < 0.001$ for
masses  $m_\nu > 45$ GeV up to $m_\nu \sim {\mathcal O}(100)$ TeV~\cite{lw}. 
Laboratory constraints for 
Dirac neutrinos are available~\cite{dir}, excluding neutrinos 
with masses between
10 GeV and 4.7 TeV. This is significant, since it precludes the possibility 
of neutrino dark matter based on an asymmetry between $\nu$ and ${\bar \nu}$~\cite{ho}. 

\subsection{Axions}

Owing to space limitations, the discussion of axions as a dark matter candidate
will be very brief.
Axions are
pseudo-Goldstone bosons which arise in solving the strong CP
problem~\cite{ax1,ax2} via a global U(1) Peccei--Quinn symmetry.  The
invisible axion~\cite{ax2} is associated with the flat direction of the
spontaneously broken PQ symmetry.  Because the PQ symmetry is also
explicitly broken (the CP violating $\theta F {\widetilde F}$ coupling is
not PQ invariant), the axion picks up a small mass similar to a pion picking
up a mass when chiral symmetry is broken.  We can expect that $m_a  \sim
m_\pi f_\pi /f_a$  where $f_a$, the axion decay constant, is the vacuum
expectation value of the PQ current and can be taken to be quite large. 
If we write the axion field as $a = f_a \theta$, near the minimum, the
potential produced by QCD instanton effects looks like $V \sim m_a^2
\theta^2 f_a^2$.  The axion equations of motion lead to
a relatively stable oscillating solution.  The energy density stored in
the oscillations exceeds the critical density~\cite{axden}  unless $f_a 
\la 10^{12}$ GeV.

	Axions may also be emitted from stars and supernovae~\cite{raff}.
In supernovae, axions are produced via nucleon--nucleon bremsstrahlung with
a coupling $g_AN \propto m_N/f_a$.  As was noted above, the cosmological
density limit requires $f_a \la 10^{12}$  GeV.  Axion emission from red
giants imply~\cite{dss}   $ f_a  \ga 10^{10}$ GeV (though this limit
depends on an adjustable axion--electron coupling), the supernova limit
requires~\cite{sn}   $ f_a \ga 2 \times 10^{11}$   GeV for naive quark
model couplings of the axion to nucleons.   
Thus only a narrow window exists for the axion as a viable dark matter
candidate.

\subsection{Neutralinos}

Supersymmetry is one of the best-motivated proposals for physics
beyond the Standard Model. It is well known that supersymmetry
could help stabilize the mass
scale of electroweak symmetry breaking by cancelling the quadratic
divergences in the radiative corrections to the mass-squared of the Higgs boson~\cite{hierarchy}. 
In addition, including supersymmetric partners of Standard
Model particles in the 
renormalization-group equations (RGEs) for the gauge couplings of the Standard Model
would permit them to unify~\cite{GUT}, 
whereas unification would not occur if only the
Standard Model particles were included in the RGEs. 

To construct the supersymmetric Standard Model~\cite{fay} we start with the
complete set of chiral fermions needed in the Standard Model, and add a scalar
superpartner to each Weyl fermion so that each field in the Standard Model
corresponds to a chiral multiplet. Similarly we must add a
gaugino for each of the gauge bosons in the Standard Model making up the
gauge multiplets. The Minimal Supersymmetric Standard Model (MSSM)~\cite{MSSM}
is defined by its minimal field content (which accounts for the known
Standard Model fields) and minimal superpotential necessary to account for
the known Yukawa mass terms. As such we define the MSSM by the superpotential
\beq
W = \epsilon_{ij} \bigl[ y_e H_1^j  L^i e^c + y_d H_1^j Q^i d^c + y_u H_2^i
Q^j u^c \bigr] + \epsilon_{ij} \mu H_1^i H_2^j .
\label{WMSSM}
\eeq 
In Eq.~(\ref{WMSSM}), the indices $\{ij\}$ are SU(2)$_L$ doublet indices.
The Yukawa couplings $y$ are all $3 \times 3$ matrices in generation space.
Note that there is no generation index for the Higgs multiplets. Colour and
generation indices have been suppressed in the above expression. There are
two Higgs doublets in the MSSM. This is a necessary addition to the Standard
Model which can be seen as arising from the holomorphic property of the
superpotential.  That is, there would be no way to account for all of the
Yukawa terms for both up-type and down-type multiplets with a single Higgs
doublet.  To avoid a massless Higgs state, a mixing term must be
added to the superpotential.

In defining the MSSM, we have limited the model to contain a minimal field
content: the only new fields are those which are {\em required} by
supersymmetry. Consequently, apart from superpartners, only the
Higgs sector was enlarged from one doublet to two. Moreover, in writing the
superpotential~(\ref{WMSSM}), we have also made a minimal choice regarding
interactions.  We have limited the types of interactions to include only
the minimal set required in the Standard Model and its supersymmetric
generalization. There are, however, additional superpotential terms which are 
consistent with gauge invariance. These would lead to rapid baryon and/or lepton number violation
and can be eliminated by imposing a
discrete symmetry on the theory called $R$-parity~\cite{Rparity}.
This can be represented as
\beq
R = (-1)^{3B + L + 2s} ,
\label{Rparity}
\eeq
where $B,L$, and $s$ are the baryon number, lepton number, and spin,
respectively. It is easy to see that, with the definition~(\ref{Rparity}), all the known
Standard Model particles have $R$-parity +1. For example, the electron has
$B=0$, $L=-1$, and $s=1/2$, and the photon has $B=L=0$ and $s=1$, so in both cases
$R=1$. Similarly, it is clear that all superpartners of the known Standard
Model particles have $R=-1$, since they must have the same value of $B$ and
$L$ as their conventional partners, but differ by 1/2 unit of spin. 
If $R$-parity is exactly conserved, then the additional
superpotential terms must be absent from the theory. 
An immediate result of imposing $R$-parity is the stability of the lightest
$R=-1$ sparticle making it a potential dark matter candidate.
Possible choices for the lightest supersymmetric particle (LSP) are
the neutralino, sneutrino, and gravitino.  Here, I will focus only on the former.

 There are four neutralinos, each of which is a  
linear combination of the $R=-1$ neutral fermions~\cite{EHNOS}: the wino
$\tilde W^3$, the partner of the third component of the $SU(2)_L$ gauge boson;
 the bino, $\tilde B$;
 and the two neutral Higgsinos,  $\tilde H_1$ and $\tilde H_2$.
The mass and composition of the LSP are determined by the gaugino masses, 
$\mu$, and  $\tan \beta$. In general,
neutralinos can  be expressed as a linear combination
\begin{equation}
	\chi = \alpha \tilde B + \beta \tilde W^3 + \gamma \tilde H_1 +
\delta 
\tilde H_2 .
\end{equation}
The solution for the coefficients $\alpha, \beta, \gamma$ and $\delta$
for neutralinos that make up the LSP
can be found by diagonalizing the mass matrix
\begin{equation}
      ({\tilde W}^3, {\tilde B}, {{\tilde H}^0}_1,{{\tilde H}^0}_2 )
  \left( \begin{array}{cccc}
M_2 & 0 & {-g_2 v_1 \over \sqrt{2}} &  {g_2 v_2 \over \sqrt{2}} \\
0 & M_1 & {g_1 v_1 \over \sqrt{2}} & {-g_1 v_2 \over \sqrt{2}} \\
{-g_2 v_1 \over \sqrt{2}} & {g_1 v_1 \over \sqrt{2}} & 0 & -\mu \\
{g_2 v_2 \over \sqrt{2}} & {-g_1 v_2 \over \sqrt{2}} & -\mu & 0 
\end{array} \right) \left( \begin{array}{c} {\tilde W}^3 \\
{\tilde B} \\ {{\tilde H}^0}_1 \\ {{\tilde H}^0}_2 \end{array} \right) ,
\end{equation}
where $M_1 (M_2)$ is a soft supersymmetry breaking
 term giving mass to the U(1) (SU(2))  gaugino(s).

The relic density of neutralinos depends on additional parameters in the MSSM beyond $M_1, M_2,
\mu$, and $\tan \beta$. These include the sfermion masses $m_{\tilde f}$ and the
Higgs pseudo-scalar mass $m_A$. To
determine the relic density it is necessary to obtain the general
annihilation cross-section for neutralinos.  In much of the parameter
space of interest, the LSP is a bino and the annihilation proceeds mainly
through sfermion exchange.

In its generality, the minimal supersymmetric standard model (MSSM) has over 100 undetermined
parameters. There are good arguments based on grand unification~\cite{GUT} and supergravity~\cite{BIM} which lead to a strong reduction in the number of parameters.
I will assume several unification conditions placed on the
supersymmetric parameters.  In all models considered, the gaugino masses
are assumed to be unified at the GUT scale with value $m_{1/2}$ 
as are the trilinear couplings with value $A_0$.  Also common to all models
considered here is the unification of all soft scalar masses set equal to $m_0$ at the GUT scale.
With this set of boundary conditions at the GUT scale, we can use the radiative electroweak 
symmetry breaking conditions by specifying the ratio of the two Higgs 
vacuum expectation values, $\tan \beta$, and the mass $M_Z$ to predict the 
values of the Higgs mixing mass parameter $\mu$ and Higgs pseudoscalar mass $m_A$.
The sign of $\mu$ remains free. 
This class of models is often referred to as the constrained MSSM (CMSSM)~\cite{funnel,cmssm,efgosi,eoss,cmssmwmap}.
In the CMSSM, the solutions for $\mu$ generally lead to a lightest neutralino
which is very nearly a pure $\tilde B$. 

\begin{figure}
 \centering \includegraphics[width=0.6\textwidth]{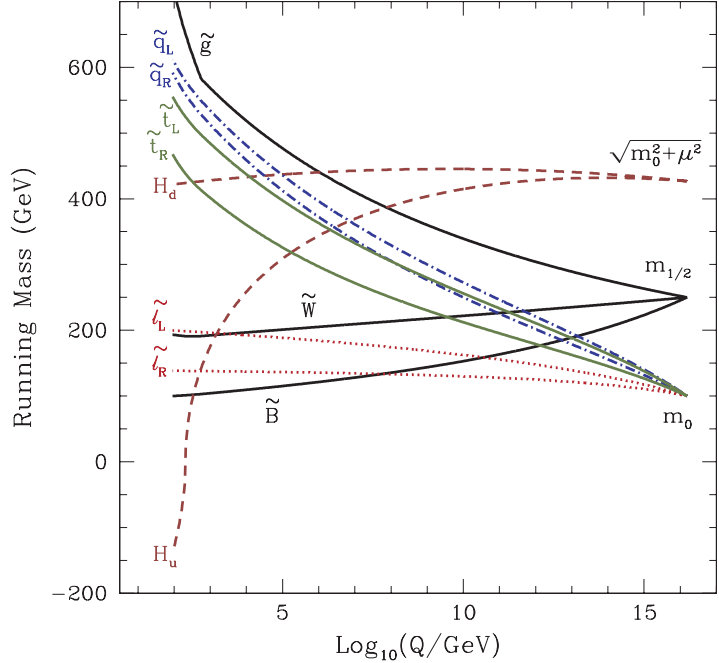}
 \vspace{-.7cm}
\caption{RG evolution of the mass parameters in the CMSSM.
I thank Toby Falk for providing this figure.}
\label{running}       
\end{figure}

In Fig.~\ref{running}, an example of the renormalization group running of the mass parameters
in the CMSSM is shown.  Here, we have chosen $m_{1/2} = 250$ GeV, $m_0 = 100$
GeV, $\tan \beta = 3$, $A_0 = 0$, and $\mu < 0$.
Indeed, it is rather amazing that, from so few input parameters, all of the
masses of the supersymmetric particles can be determined. 
The characteristic features that one sees in the figure are, for example, that
the coloured sparticles are typically the heaviest in the spectrum.  This is
due to the large positive correction to the masses due to $\alpha_3$ in the
RGEs.  Also, one finds that the $\widetilde{B}$ is typically the lightest
sparticle.  But most importantly, notice that one of the Higgs mass$^2$ goes
negative triggering electroweak symmetry breaking~\cite{rewsb}. (The negative
sign in the figure refers to the sign of the mass squared, even though it is the
mass of the sparticles which is depicted.) 

\begin{figure}
 \centering  \includegraphics[width=0.45\textwidth]{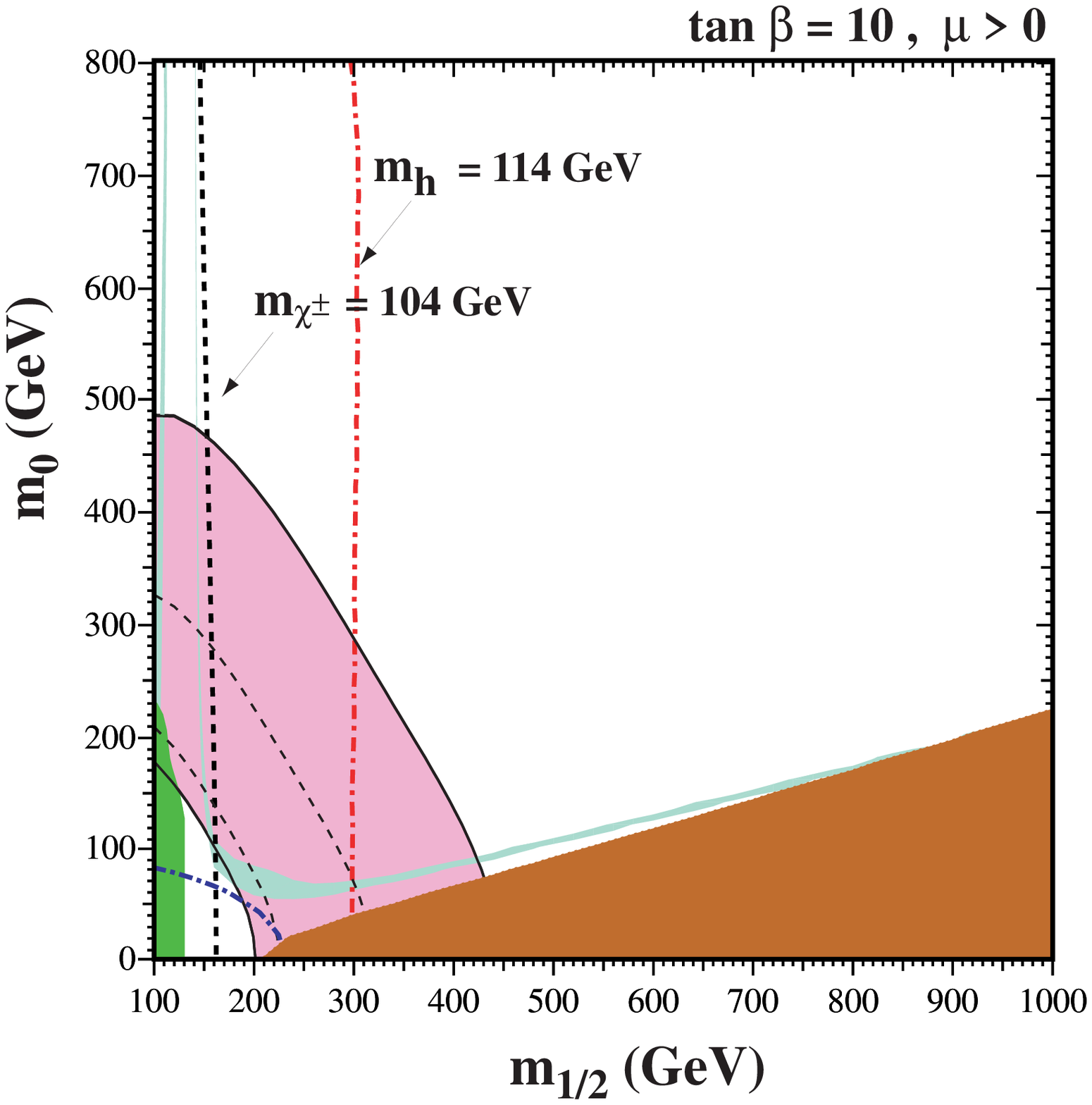} \includegraphics[width=0.45\textwidth]{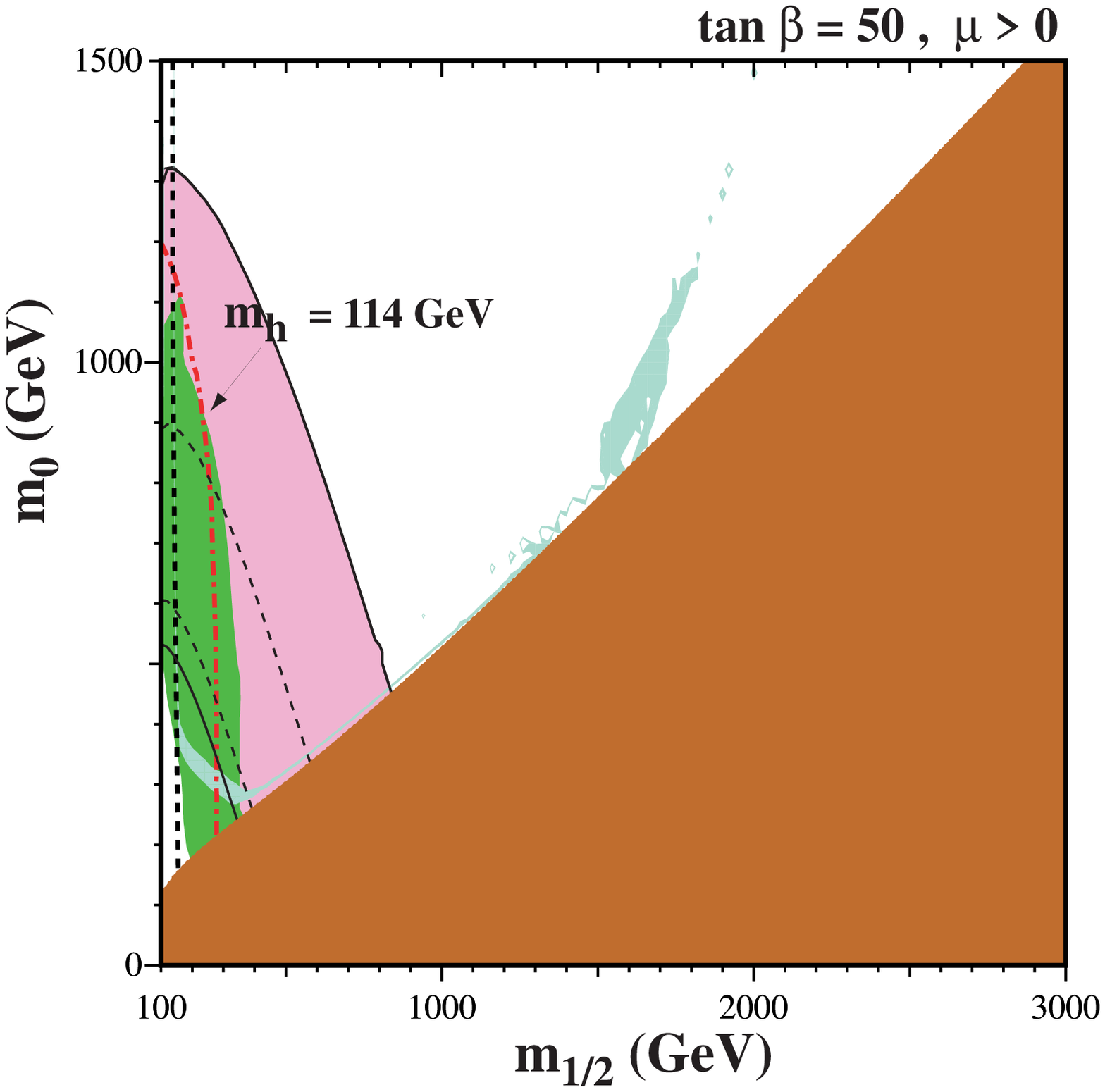}
\caption{The $(m_{1/2}, m_0)$ planes for  (a) $\tan \beta = 10$ and  $\mu > 0$, 
assuming $A_0 = 0, m_t = 175$~GeV and
$m_b(m_b)^{\overline {MS}}_{SM} = 4.25$~GeV. The near-vertical (red)
dot-dashed lines are the contours $m_h = 114$~GeV, and the near-vertical (black) dashed
line is the contour $m_{\chi^\pm} = 104$~GeV. Also
shown by the dot-dashed curve in the lower left is the corner
excluded by the LEP bound of $m_{\tilde e} > 99$ GeV. The medium (dark
green) shaded region is excluded by $b \to s
\gamma$, and the light (turquoise) shaded area is the cosmologically
preferred region. In the dark
(brick red) shaded region, the LSP is the charged ${\tilde \tau}_1$. The
region allowed by the E821 measurement of $a_\mu$ at the 2-$\sigma$
level, is shaded (pink) and bounded by solid black lines, with dashed
lines indicating the 1-$\sigma$ ranges. In (b), $\tan \beta= 50$. }
\label{fig:UHM}       
\end{figure}

For a given value of $\tan \beta$, $A_0$,  and $sgn(\mu)$, the resulting regions of 
parameter space with
acceptable relic density and which satisfy the phenomenological constraints
can be displayed on the  $m_{1/2} - m_0$ plane.
In Fig.~\ref{fig:UHM}(a),  the light
shaded region corresponds to that portion of the CMSSM plane
with $\tan \beta = 10$, $A_0 = 0$, and $\mu > 0$ such that the computed
relic density yields the WMAP value given in Eq.~(\ref{wmap})~\cite{eoss}.
The bulk region at relatively low values of 
$m_{1/2}$ and $m_0$,  tapers off
as $m_{1/2}$ is increased.  At higher values of $m_0$,  annihilation cross sections
are too small to maintain an acceptable relic density and $\Omega_\chi h^2$ is too large.
Although sfermion masses are also enhanced at large $m_{1/2}$ (due to RGE running),
co-annihilation processes between the LSP and the next lightest sparticle 
(in this case the $\tilde \tau$) enhance the annihilation cross section and reduce the
relic density.  This occurs when the LSP and NLSP are nearly degenerate in mass.
The dark shaded region has $m_{\tilde \tau}< m_\chi$
and is excluded.   The effect of co-annihilations is
to create an allowed band about 25--50 GeV wide in $m_0$ for $m_{1/2} \la
950$ GeV, or $m_{1/2} \la 400$ GeV, which tracks above the $m_{{\tilde \tau}_1}
 =m_\chi$ contour~\cite{efo}.  

Also shown in Fig.~\ref{fig:UHM}(a) are
the relevant phenomenological constraints.  
These include the LEP limits on the chargino mass: $m_{\chi^\pm} > 104$~GeV~\cite{LEPsusy};
on the selectron mass: $m_{\tilde e} > 99$~GeV~ \cite{LEPSUSYWG_0101};
and on the Higgs mass: $m_h >
114$~GeV~\cite{LEPHiggs}. The former two constrain $m_{1/2}$ and $m_0$ directly
via the sparticle masses, and the latter indirectly via the sensitivity of
radiative corrections to the Higgs mass to the sparticle masses,
principally $m_{\tilde t, \tilde b}$. 
{\tt FeynHiggs}~\cite{FeynHiggs} is used for the calculation of $m_h$. 
The Higgs limit  imposes important constraints
principally on $m_{1/2}$ particularly at low $\tan \beta$.
Another constraint is the requirement that
the branching ratio for $b \rightarrow
s \gamma$ be consistent with the experimental measurements~\cite{bsgex}. 
These measurements agree with the Standard Model, and
therefore provide bounds on MSSM particles~\cite{gam}  such as the chargino and
charged Higgs masses, in particular. Typically, the $b\rightarrow s\gamma$
constraint is more important for $\mu < 0$, but it is also relevant for
$\mu > 0$,  particularly when $\tan\beta$ is large. The constraint imposed by
measurements of $b\rightarrow s\gamma$ also excludes small
values of $m_{1/2}$. 
Finally, there are
regions of the $(m_{1/2}, m_0)$ plane that are favoured by
the BNL measurement~\cite{newBNL} of $g_\mu - 2$ at the 2-$\sigma$ level, corresponding to 
a deviation  from the Standard Model 
calculation~\cite{Davier}.

Another
mechanism for extending the allowed regions in the CMSSM to large
$m_\chi$ is rapid annihilation via a direct-channel pole when $m_\chi
\sim {1\over 2} m_{A}$~\cite{funnel,efgosi}. Since the heavy scalar and
pseudoscalar Higgs masses decrease as  
$\tan \beta$ increases, eventually  $ 2 m_\chi \simeq  m_A$ yielding a
`funnel' extending to large
$m_{1/2}$ and
$m_0$ at large
$\tan\beta$, as seen in Fig.~\ref{fig:UHM}(b).
As one can see, the impact of the Higgs mass constraint is reduced (relative to 
the case with $\tan \beta = 10$) while that of $b \to s \gamma$ is enhanced.

\begin{figure}
\centering\includegraphics[width=0.6\textwidth]{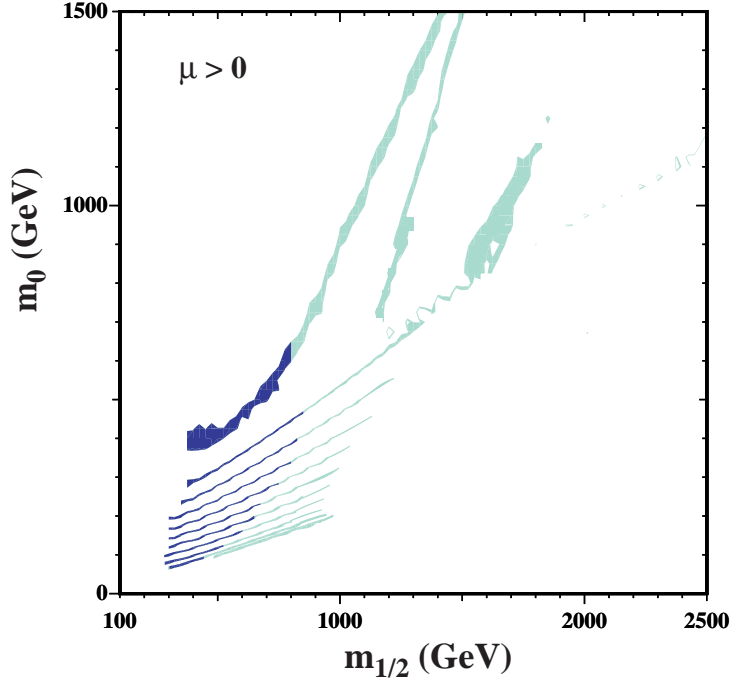}
\caption{
The strips display the regions of the $(m_{1/2}, m_0)$ plane that are
compatible with the WMAP determination of $\Omega_\chi h^2$ and the laboratory
constraints for $\mu > 0$ and $\tan \beta = 5, 10, 15, 20, 25, 30,
35, 40, 45, 50, 55$. The parts of the strips compatible with $g_\mu - 2$ 
at the 2-$\sigma$ level have darker shading.}
\label{fig:strips}
\end{figure}

Shown in Fig.~\ref{fig:strips} are the WMAP lines~\cite{eoss} of the $(m_{1/2}, m_0)$
plane for $\mu > 0$ and
values of $\tan \beta$ from 5 to 55, in steps $\Delta ( \tan \beta ) = 5$.
We notice immediately that the strips are considerably narrower than the
spacing between them, though any intermediate point in the $(m_{1/2},
m_0)$ plane would be compatible with some intermediate value of $\tan
\beta$. The right (left) ends of the strips correspond to the maximal
(minimal) allowed values of $m_{1/2}$ and hence $m_\chi$. 
The lower bounds on $m_{1/2}$ are due to the Higgs 
mass constraint for $\tan \beta \le 23$, but are determined by the $b \to 
s \gamma$ constraint for higher values of $\tan \beta$.

\begin{figure}
\begin{center}
\includegraphics[width=0.5\textwidth]{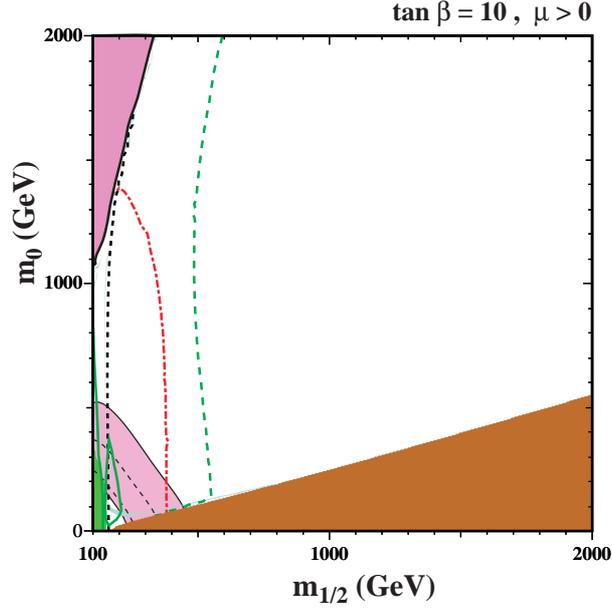}
\end{center}
\caption{
As in Fig.~\ref{fig:UHM}, showing the $m_{1/2},m_0$ plane extended to high values of $m_0$}
\label{fig:fp}
\end{figure}

Finally, there is one additional region of acceptable relic density known as the
focus-point region~\cite{fp}, which is found
at very high values of $m_0$. 
An example showing this region is found in Fig.~\ref{fig:fp},
plotted for $\tan \beta = 10$, $\mu > 0$, and $m_t = 172.4$ GeV.
As $m_0$ is increased, the solution for $\mu$ at low energies as determined
by the electroweak symmetry breaking conditions eventually begins to drop. 
When $\mu \la m_{1/2}$, the composition of the LSP gains a strong Higgsino
component and as such the relic density begins to drop precipitously. 
As $m_0$ is increased further, there are no longer any solutions for $\mu$.  This 
occurs in the shaded region in the upper left corner of Fig.~\ref{fig:fp}. 
The position of the focus point strip is very sensitive to the value of $m_t$~\cite{fine}.

As seen in Fig.~\ref{fig:UHM}, the relic density constraint is compatible
with relatively large values of $m_{1/2}$ and $m_0$. However, all
values of $m_{1/2}$ and $m_0$ are not equally viable when 
the available phenomenological and cosmological constraints are taken into account.
A global likelihood analysis enables one to 
pin down the available parameter space in the CMSSM. 
One can avoid the dependence on priors by performing
a pure likelihood analysis as in Ref.~\cite{Ellis:2003si}, or a purely $\chi^2$-based fit as 
done in Refs.~\cite{Ellis:2007fu,Buchmueller:2007zk}.  
Here we present results from one such analysis~\cite{Buchmueller:2008qe,Buchmueller:2009fn}, 
which used a Markov--Chain
Monte Carlo (MCMC) technique to explore efficiently the likelihood function in
the parameter space of the CMSSM. A full list of the observables and the values assumed
for them in this global analysis are given in Ref.~\cite{Buchmueller:2007zk}, and updated 
in Refs.~\cite{Buchmueller:2008qe,Buchmueller:2009fn}. 

\begin{figure}[ht]
\vskip .2in
\centering \begin{picture}(300,200)
  \put( 20,   20){ \resizebox{0.65\textwidth}{!}{\includegraphics{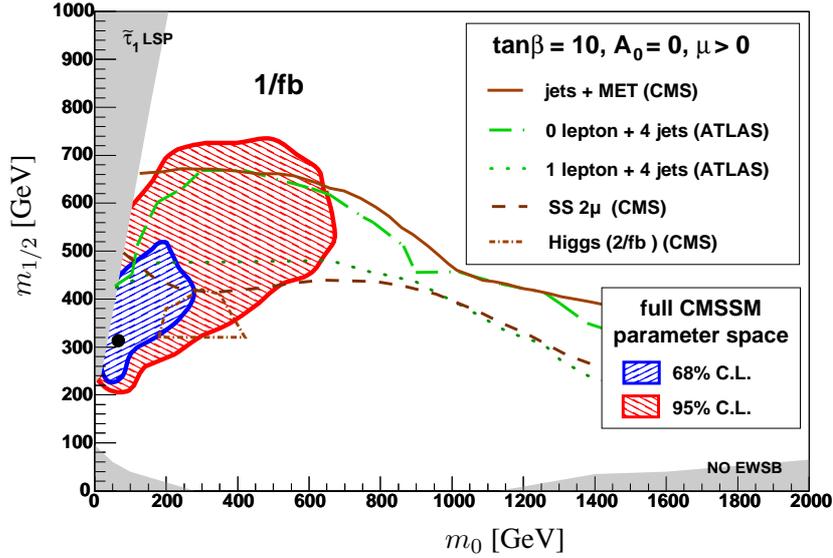}}}
  \put(170,   5){$m_0$ [GeV]}
  \put( 15,   100){\begin{rotate}{90}$m_{1/2}$ [GeV]\end{rotate}}
\end{picture}
\caption{\label{fig:MCMC} The $(m_0, m_{1/2})$ plane in the CMSSM
showing the regions favoured in a likelihood analysis
at the 68\% (blue) and 95\% (red) confidence levels~\protect\cite{Buchmueller:2008qe}. The best-fit
point is shown by the black point. Also shown are
the discovery contours in different channels for the LHC with 1/fb
(2/fb for the Higgs search in cascade decays of sparticles)~\protect\cite{:1999fr,Ball:2007zza}.}
\end{figure}

The 68\% and 95\% confidence-level (CL) regions in the
$(m_{1/2}, m_0)$ plane of the CMSSM is shown in
Fig.~\ref{fig:MCMC}~\cite{Buchmueller:2008qe}. Also shown for comparison are the physics
reaches of ATLAS and CMS with 1/fb of integrated luminosity~\cite{:1999fr,Ball:2007zza}. 
(MET stands for missing
transverse energy, SS stands for same-sign dilepton pairs, and the
sensitivity for finding the lightest Higgs boson in cascade decays of
supersymmetric particles is calculated for 2/fb of data.) The likelihood
analysis assumed $\mu > 0$, as motivated by the sign of the
apparent discrepancy in $g_\mu - 2$, but sampled all values of $\tan \beta$
and $A_0$: the experimental sensitivities were estimated
assuming $\tan \beta = 10$ and $A_0 = 0$, but are probably not
very sensitive to these assumptions. The global maximum of the
likelihood function (indicated by the black dot) is at
$m_{1/2} = 310$~GeV,
$m_0 = 60$~GeV, $A_0 = 240$~GeV, $\tan \beta = 11$, and
$\chi^2/N_{dof} = 20.4/19$ (37\% probability). It is encouraging that the best-fit point lies
well within the LHC discovery range, as do the 68\% and most of the
95\% CL regions. 

Improvements in sensitivity have made it possible for direct detection experiments~\cite{cdms,xenon10}
to be competitive with other phenomenological constraints.
The elastic cross section for $\chi$ scattering on a nucleus can be
decomposed into a scalar (spin-independent)  and a spin-dependent
part.  Each of these can be
written in terms of the cross sections for elastic scattering off individual nucleons.
The scalar part of the cross section can be
written as
\begin{equation} \label{eqn:sigmaSI}
  \sigma_{\rm SI} = \frac{4 m_{r}^{2}}{\pi}
                    \left[ Z f_{p} + (A-Z) f_{n}  \right]^{2},
\end{equation}
where $m_r$ is the $\chi$-nuclear reduced mass and
\begin{equation} \label{eqn:fN}
  \frac{f_N}{m_N}
    = \sum_{q=u,d,s} \fNTq{q} \frac{\alpha_{3q}}{m_{q}}
      + \frac{2}{27} f_{TG}^{(N)}
        \sum_{q=c,b,t} \frac{\alpha_{3q}}{m_q} ,
\end{equation}
for $N$ = p or n.  The parameters $\fNTq{q}$ are defined by
\begin{equation} \label{eqn:Bq}
  m_N \fNTq{q}
  \equiv \langle N | m_{q} \bar{q} q | N \rangle
  \equiv m_q \BNq{q} ,
\end{equation}
and the $\alpha_{3q}$ contain the individual quark-neutralino scattering cross sections,
see Refs.~\cite{EFlO1,eoss8,eosv} for further details regarding the calculation of the cross section.

The elastic scattering of neutralinos on nucleons is very sensitive to the strangeness
contribution to the nucleon mass and can be characterized by the parameter $y$ 
which is also related to the $\pi$-nucleon sigma term $\SigmapiN$ by
\begin{equation} \label{eqn:y2}
  y \equiv {2B_s \over B_u + B_d} = 1 - \sigma_0/\SigmapiN \; ,
\end{equation}
where $\sigma_0$ is the change in the nucleon mass due to non-zero $u$ and $d$ masses and is estimated from octet baryon mass differences to be 
$\sigma_0 = 36$~MeV~\cite{Borasoy:1996bx},
and the latest determination of $\SigmapiN = 64$~MeV.
The effects of varying these assumptions are discussed in the context of the CMSSM
in Refs.~\cite{eoss8,eosv}. Lattice calculations are now reaching the stage where they may also provide useful information on $\Sigma_{\pi N}$~\cite{0901.3310}, and a recent analysis
would suggest a lower value $\Sigma_{\pi N} \la 40$~\cite{joel}.

\begin{figure}[ht!]
\centering \includegraphics[width=0.4\textwidth]{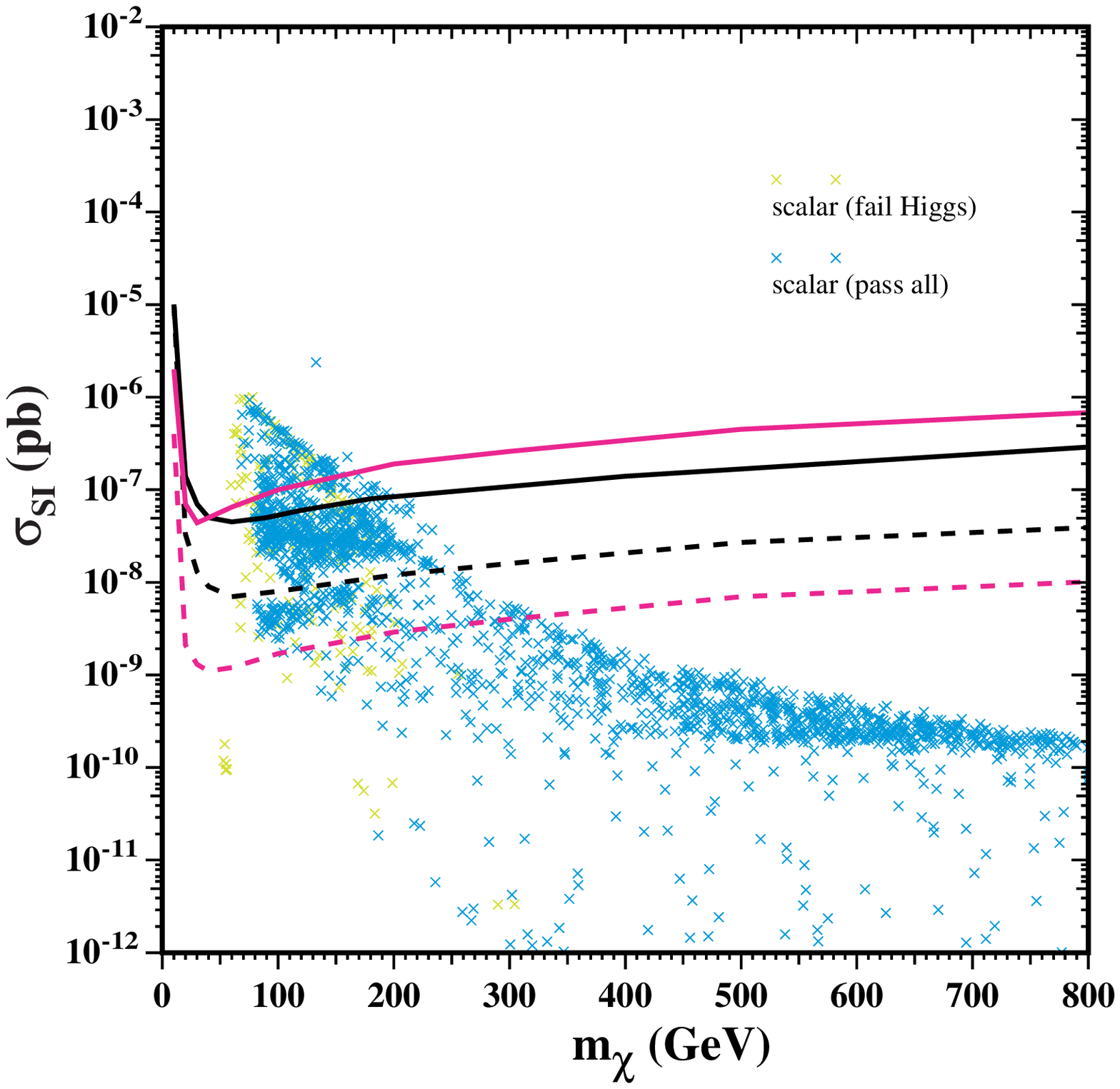} \includegraphics[width=0.55\textwidth]{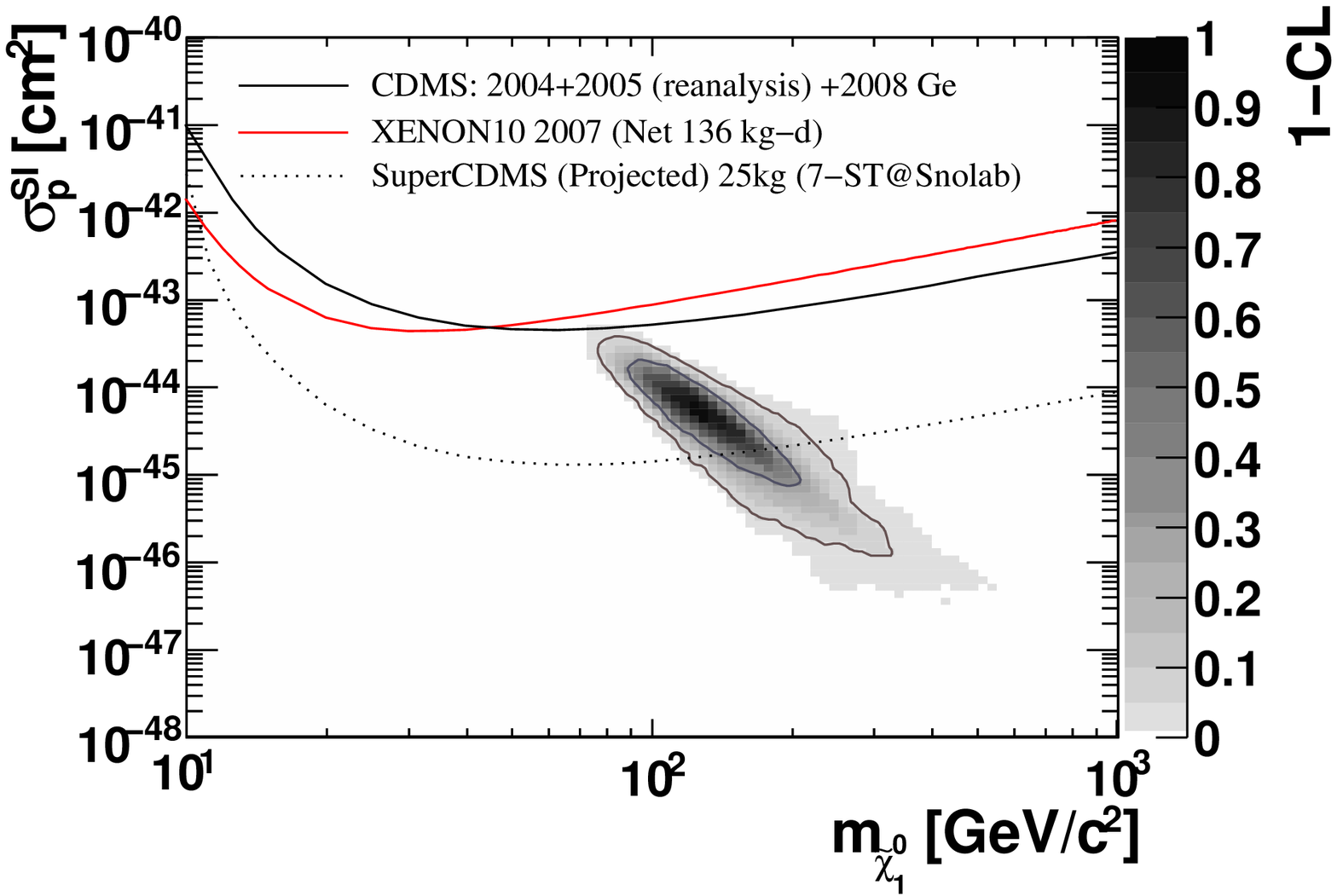}
\caption{(a) The entire potential range of neutralino--nucleon cross sections as functions of neutralino mass for the CMSSM, with $5 \leq \tan \beta \leq 55$, 0 $\leq m_{1/2} \leq 2000$ GeV, 100 GeV $\leq m_0 \leq 2000$ GeV, and $-3 m_{1/2} \leq A_0 \leq 3 m_{1/2}$.   Also shown are upper limits on the spin-independent 
dark matter scattering cross section from CDMS~II~\protect\cite{cdms} (solid black line) and
XENON10~\cite{xenon10} (solid pink line), as well as the expected sensitivities  
for XENON100~\cite{XENON100} (dashed pink line) and SuperCDMS at the Soudan 
Mine~\protect\cite{superCDMS} (dashed black line). Taken from Ref.~\protect\cite{eosk5}.
(b) The 68\% and 95\% CL areas in the cross section--mass plane from the frequentist analysis
of Ref.~\protect\cite{Buchmueller:2009fn}.
\label{fig:cmssm}}
\end{figure}

In Fig.~\ref{fig:cmssm}(a) we show CMSSM spin-independent
neutralino-nucleon cross section,  as obtained in a scan over all 
CMSSM parameters with $5 \leq \tan \beta \leq 55$, 100 $\leq m_{1/2} \leq 2000$~GeV, 
0 GeV~$\leq m_0 \leq 2000$~GeV, and $-3 m_{1/2} \leq A_0 \leq 3 m_{1/2}$~\cite{eosk5}.  
We also allow both positive and negative $\mu$, except for large $\tan \beta > 30$, 
where convergence becomes difficult in the $\mu < 0$ case.  
 At low $m_{\chi} < 300$~GeV, cross sections 
generally exceed $10^{-9}$~pb, and the largest scalar cross sections, 
which occur for $m_{\chi} \sim 100$ GeV, are already excluded by CDMS~II~\cite{cdms} and/or 
XENON10~\cite{xenon10}. 
These exclusions occur primarily in the focus-point region at large $\tan \beta$.  
On the other hand, for $m_{\chi} \ga 400$~GeV, scalar cross sections are well 
below $10^{-9}$ pb, and come from the co-annihilation strip or the rapid-annihilation 
funnel that appears at large $\tan \beta$ in the CMSSM. The effective cross sections shown are
suppressed for points with $\Omega_\chi \ll \Omega_{CDM}$, and there may be cancellations
at larger $m_{\chi}$ that suppress the cross sections substantially. 
These regions of parameter space will not be probed by direct detection experiments in the near future~\cite{XENON100,superCDMS}.   The corresponding 68\% and 95\% CL regions in
the cross section--neutralino mass plane from the frequentist analysis of Ref.~\cite{Buchmueller:2009fn} are shown in Fig.~\ref{fig:cmssm}(b).

\section*{Acknowledgements}

This work was supported in part by DOE grant DE-FG02-94ER-40823

\end{document}